\documentstyle[preprint,epsfig,eqsecnum,aps]{revtex}

\newcommand{\be}{\begin{equation}}
\newcommand{\ee}{\end{equation}}
\newcommand{\bea}{\begin{eqnarray}}
\newcommand{\eea}{\end{eqnarray}}
\begin{document}
\draft
\title{New results for $2\nu\beta\beta$ decay with large particle-particle 
two body proton-neutron interaction}
\author{A. A. Raduta$^{a,b,c)}$, O. Haug $^{c)}$, F. \v {S}imkovic$^{c,d)}$ 
and Amand Faessler$^{c)}$ }
\address{$^{a)}$Institute of Physics and Nuclear Engineering, Bucharest, 
POB MG6, Romania}
\address{$^{b)}$Dep. of Theoretical Physics and
Mathematics, Bucharest University, POB MG11, Romania}
\address{$^{c)}$Institut f\"ur Theoretische Physik,
Auf der Morgenstelle 14, T\"uebingen, Germany}
\address{$^{d)}$Comenius University, Mlynska dolina F1, Bratislava, Slovakia}
\date{\today}
\maketitle
\begin{abstract}
A model many-body Hamiltonian describing an heterogenous system of paired 
protons and paired neutrons and interacting among themselves through
monopole particle-hole and monopole particle-particle interactions is used to
study the double beta decay of Fermi type. The states are described by time 
dependent approaches choosing as trial functions coherent states of the 
symmetry groups underlying the model Hamiltonian. One formalism, VP1, is 
fully equivalent with the standard pnQRPA and therefore fails at a critical 
value of the particle-particle interaction strength while another one, 
VP2, corresponds to a two step BCS treatment, i.e. the proton quasiparticles 
are paired with the neutron quasiparticles. In this way a harmonic 
description for the double beta transition amplitude is provided for any
strength of the particle-particle interaction. The approximation quality is
judged by comparing the actual results with the exact result as well as 
with those corresponding to various truncations of the boson expanded 
Hamiltonian and transition operator. Finally it is shown that the dynamic 
ground states provided by VP1 and VP2 are reasonable well approximated by 
solutions of a variational principle. This remark constitutes a step forward 
finding an 
approach where the RPA ground state is a solution of a variational 
principle equation. 
\end{abstract}
\section{Introduction}
\label{sec: level1}
One of the most exciting subject of the modern nuclear physics
is the double beta decay. The reason is that from this field
one expects an answer to the question of whether neutrino is a
Majorana or a Dirac particle. The process may take place through
one of two channels, 0$\nu\beta\beta$ and $2\nu\beta\beta$. The
answer to the above mentioned question might come from the
discovery of the first process which if it exists requires that the neutrino
is a Majorana particle and which should be a
rare one \cite{Hax,Doi,Fas,Tom,SuCi,Fas1}.
In order to make predictions of the neutrino mass and the right
handedness of its electroweak interactions one should use
reliable nuclear matrix elements. However there is no stringent
test for these matrix elements. Fortunately for the double beta
decay with two anti-neutrinos in the final state $2\nu\beta\beta$, 
one uses similar
nuclear matrix elements and moreover for this process plenty of
data exist. Therefore the idea of using for the neutrino-less
double beta decay the same many-body approach and the same NN force
which describe
realistically the two neutrino double beta decay, was adopted by
most of the groups involved in such studies. Actually this is
the reason why so many theoreticians focused their attention in
explaining the features of the experimentally 
measured $2\nu\beta\beta$ process 
\cite{vogel1,vogel2,civ1,grotz,Klap,mut,sutai}.
The formalism which predicts decay rates closest to the
experimental data is the proton-neutron quasiparticle random
phase approximation (pnQRPA) which includes the 
particle-particle ($pp$) two body interaction. Without this interaction the
predicted transition amplitude is too large as compared to the
existent data. Including the $pp$ interaction and considering
its strength, $g_{pp}$, as a free parameter one obtains the following
behavior for the transition amplitude. For a large interval of
$g_{pp}$, starting from zero, there exists a plateau followed by a
quick decrease. The amplitude vanishes at a certain $g_{pp}$ and
shortly after this point one reaches, with increasing $g_{pp}$,
 the critical value where
the pnQRPA breaks down. On the decreasing part
of the $2\nu\beta\beta$ transition amplitude as a function of $g_{pp}$
one finds agreement with experiment.
The drawback of this description
is that in this area of $g_{pp}$, the pnQRPA is not a good
description and therefore the results are not stable to adding
higher RPA correlations. Many attempts have been made to
stabilize the ground state in this region of the particle-particle interaction.
The new methods tried either to keep the pnQRPA boson picture
but include higher correlations through boson expansion
\cite{rad1,rad2,suh1,gri,RadSuh1,RadSuh2}
techniques or via self-consistent procedures \cite{SimSmo}, 
or to re-normalize the pnQRPA phonon
\cite{Toi95,Schw96,SimKam,RadRad}.  These methods
improved the description of the data\cite{Rasifa} although they have 
also drawbacks and
lack sometimes of consistency \cite{RadRad,RaHa,RaPa}. 

In a previous paper \cite{RaHa} we advanced the idea that the ground state
may be stabilized by introducing the particle-particle interaction first in
the mean field and then to the pnQRPA process. A  way to
do that is even possible at the level of independent quasiparticle
representation. Indeed, we noticed in a schematic single j model \cite{Samb}
that one term of the two body quasiparticle interaction is a
quasiparticle proton-neutron pairing interaction and its
strength is negative.
Therefore one could define a Cooper pair out of one proton
quasiparticle and
one neutron quasiparticle. This was worked out through 
a time dependent variational principle. We chose alternatively
four distinct trial functions. Two of these define 
different RPA approaches. One is identical to the standard
pnQRPA and works for small strengths of the particle-particle interaction,
namely before the critical point, whereas the other one provides
a harmonic picture beyond the critical value of the interaction
strength, i.e. in the region where the standard pnQRPA does not
work at all. Moreover in this region a boson expansion procedure
might be defined and through diagonalization the exact result
can be recovered.

In the present paper we continue the study started in the
previous paper, by focusing on the following new features.
Here we calculate the double beta decay transition amplitude by
using the description in the interval of the particle-particle interaction
forbidden for the pnQRPA approach. We hope that by matching the
approaches in the two complementary intervals one could provide
a unified description of the process for small as well for large
strengths of the particle-particle interaction.
When one uses the boson expansion for the model Hamiltonian one
aims at testing the convergence properties by comparing the
corresponding results with the exact ones. One should mention
that this is possible only for a single j  model Hamiltonian
involving a monopole two body interactions.
Formulating it for the double beta decay Fermi transition the boson
expansion of the transition operator emitted in ref.\cite{rad1} is a
Schwinger type boson expansion since it uses two different
bosons in order to satisfy the condition of mapping two algebras.
This is a good test for the boson representation which was
previously used. 
Another aspect which is treated here refers to the structure of
the vacuum states provided by the standard pnQRPA and the
approach which works beyond the critical particle-particle interaction
 strength. We want
to check whether these two vacua might be described in a
reasonable approximative fashion by static ground states yielded
by a variational principle with a suitable trial function.

This project is achieved according to the following plan.
In Section 2, a brief review of the results obtained in the
previous paper is given. These results are used here for a self-consistent
presentation. 
In Section 3, we compare the results for double beta decay
transition amplitude obtained in the two intervals  of the 
particle-particle interaction, i.e. before and after its critical value, with
different methods.
In Section 4 the Holstein-Primakoff boson representation is used
for the model Hamiltonian. By diagonalizing the full
Hamiltonian, the first order approach and the results for the limit
$\Omega\rightarrow\infty$ and by comparing the results with the exact
ones and those obtained within the harmonic picture, one could
judge the convergence quality of the expansion used.
The Schwinger type boson representation is here also applied.
The possible contribution of pairing vibrational states to the
enhancement of the decay rate is discussed. The decay to the
double phonon pairing vibration is explicitly analyzed.
In Section 5 we study a possible relationship between the vacuum
states of the standard pnQRPA, the second order BCS approach,
and the static ground state provided by classical variational
equations obtained with two different trial functions.
Final conclusions are drawn in Section 6.
\section{Brief review of some previous results}
\label{sec:  level2}
Since the present paper is continuing the study we started in
previous publications we give here a brief summary of the
main ingredients used there \cite{RaHa}. In this way we fix the notation
and the conventions. Also the general frame of our present
results will be better emphasized.
The object of our study is a system of protons and neutrons
moving in a spherical shell model mean field, interacting
among themselves through pairing, particle-hole and
particle-particle monopole two body interaction. The associated
Hamiltonian reads:
\bea
H&=&\epsilon_p\sum_{m}c^{\dagger}_{pm}c_{pm}+\epsilon_n\sum_{m}
c^{\dagger}_{nm}c_{nm}
\nonumber \\
&-&\frac{G_p}{2}P^{\dagger}_pP_p-\frac{G_n}{2}P^{\dagger}_nP_n
\nonumber \\
&+&2\chi \beta^-\beta^+-2\chi_1 P^-P^+~~,
\eea
where the following notations have been used:
\bea
P^{\dagger}_p&=&\sum_{m}c^{\dagger}_{\tau
m}c^{\dagger}_{\widetilde{\tau m}}, \tau=p,n,
\nonumber\\
\beta^-  &=&\sum_{m}c^{\dag}_{pm}c_{nm}; \beta^+=(\beta^-)^+,
\nonumber\\
P^- &=&\sum_{m}c^{\dag}_{pm}\widetilde {c}^{\dag}_{nm};
P^+=(P^-)^+.
\eea
The standard notations for the creation and annihilation
operators are used. For the sake of simplicity we consider here
the case of a single j shell. The extension to the multi-shell
picture is straightforward. However for the purposes of the
present study the multi-shell calculations are not necessary. The
shell j is self-understood in our  notations. To specify
the isospin of the particle occupying the shell we use the
indices p for proton and n for neutrons. The time reversed state
is denoted by $|\widetilde{\tau m}\rangle=(-1)^{j-m}|\tau,-m\rangle$.
Performing the Bogoliubov-Valatin transformation, the model
Hamiltonian is transformed in a many body quasiparticle operator
which after ignoring the quasiparticle scattering terms and, for
the sake of simplicity, equating the proton and neutron single
particle energies, has the form:
\be
H=\epsilon (\hat{N}_p+\hat{N}_n)+\lambda_1A^{\dag}_{pn}A_{pn}+\lambda_2
(A^{\dag}_{pn}A^{\dag}_{pn}+A_{pn}A_{pn}),
\ee
where ${\hat N}_p$ and ${\hat N}_n$ are the proton and neutron
quasiparticle number operators, respectively, while $A^+_{pn},A_{pn}$ stand
for the proton-neutron pairing operators built up with the
quasiparticle operators $a^{\dag}_{pm},a^{\dag}_{nm}$:
\be
A^{\dag}_{pn}=[a^{\dag}_pa^{\dag}_n]_{00},\; A_{pn}=(A^{\dag}_{pn})^{\dag}.
\ee
The coefficients $\epsilon,\lambda_1,\lambda_2$ are functions of
the parameters defining the Hamiltonian (2.1) as well as of the
coefficients U, V determining the quasiparticle representation.
The Hamiltonian  (2.3) is a quadratic expression in the generators of
the SU(2) algebra
\be
\hat{S}_0=\frac{1}{2}[\hat {N}_p+\hat {N}_n-2\Omega],~~
\hat {S}_+=\sqrt{2\Omega}A^{\dag}_{pn},~~
\hat {S}_-=\sqrt{2\Omega}A_{pn},~~\Omega=\frac{2j+1}{2}.
\ee
and therefore it is exactly solvable. This feature has the great
advantage that one could test the many body approximations
adopted in various formalisms. This virtue made it very
attractive for many theoreticians who used it widely to advance
some new hypothesis for treating the proton-neutron two body
interaction \cite{Samb,Civ1,Civ2}.  

One may argue that restricting the model space to a single j level one misses
the effect of spin-orbit interaction. However from our 
earlier studies \cite{Rad11,Rad12},
we now that the spin orbit coupling is important for the Gamow-Teller double
beta transition, where most of the strength is carried by 
spin-flip configurations,
 but not for the Fermi transition which we are studying
in the present paper.

If one assumes that the monopole two quasiparticle
operators and their hermitian conjugate satisfy bosonic like
commutation relations than a harmonic excitation operator might
be defined. Therefore the Weyl algebra, defined by the
operators $A^{\dagger},A, I$ (here $I$ denotes the unity
operator), can be used to describe the eigenstates of H.  When
the terms with four quasiparticles in H are dominant and
the quasiboson approximation is adopted, for the above
mentioned commutators, one could diagonalize first the anharmonic
part of H and then treat the term expressing the quasiparticle total
number operator. Up to an additive constant, the
anharmonic part of H (four quasiparticle terms) is a linear
combination of the generators of the group $SU(1,1)$, i.e.
${A^+}^2,A^2,2+4A^+A$. Therefore this symmetry might be useful
for describing the eigenstates of H.

The model Hamiltonian (2.1) was treated, in ref.\cite{RaHa},
\footnote{Throughout this paper we adopt units where $\hbar$=1} 
within a dependent variational principle formalism:
\be
\delta \int^{t}_{0}\langle \psi|H-i\frac {\partial}{\partial t^\prime}|\psi
\rangle dt^{\prime}=0.
\ee
by taking as trial functions the coherent states of the above
mentioned symmetry groups.  Since the coherent state for SU(2)
and for the Weyl group are formally identical with the difference that
in the first case the operators $A^{\dagger}, A$ satisfy exact
commutation relations, while in the later case they are
considered as quasibosons, we complete our study with a fourth
wave function which is similar to the SU(1,1) coherent state
but the operators involved satisfy exact commutation relations.
Concluding, the following trial functions have been
alternatively considered for the variational state \cite{RaHa,RaHa2}:
\bea
|\psi_1(z,z^*)\rangle&=&exp\{zA^{\dag}_{pn}-z^*A_{pn}\}|0\rangle_q,~
\left[A_{pn},A^{\dagger}_{pn}\right]=1,~~\rm{(VP1)}
\\
|\psi_2(z,z^*)\rangle&=&exp\{zA^{\dag}_{pn}-z^*A_{pn}\}|0\rangle_q,
\left[A_{pn},A^{\dagger}_{pn}\right]=1-\frac{1}{2\Omega}({\hat
N}_p+{\hat N}_n), ~~\rm{(VP2)}\\
|\psi_3(z,z^*)\rangle&=&\exp(z{A^{\dag}_{pn}}^2-z^*{A_{pn}}^2)
|0\rangle_q,~
\left[A_{pn}, A^{\dag}_{pn}\right]=1,~~\rm{(VP3)}
 \\
|\psi_4(z,z^*)\rangle &=&{\cal
N}_4e^{z{A^{\dag}_{pn}}^2}|0\rangle_q,~
[A_{pn},A^{\dagger}_{pn}]=1-\frac{1}{2\Omega}({\hat N}_p+{\hat N}_n),
~~\rm{(VP4)}
\eea
In each case, $z$ is a complex function of time and $z^*$
denotes its complex conjugate. $|0\rangle_q$ is the
quasiparticle vacuum state.  The formalism corresponding to the
trial function $|\Psi_k\rangle$ with the commutation relations
specified above, will be hereafter called VPk.  For each of the four
cases we solved the static equations of motion and then found
the harmonic mode describing small oscillations around the
static ground state. Several properties such as the behavior of
energies with respect to the strength of the particle-particle
interaction, as the ground state correlations, as the single beta
transition amplitude, as the Ikeda sum rule and also  various quantization
procedures of the classical equations of motion have been
analyzed.
 For a better presentation the
final results will be considered, as in ref.\cite{RaHa,Samb}, 
as function of the re-scaled
strengths of particle-hole and particle-particle
interactions:
\be
\chi^{\prime}=2\Omega \chi,~~k^{\prime}=2\Omega \chi_1.
\ee
For example the energies of the first excited states in mother
and daughter nuclei, predicted by VP1 and VP2, vary with $k'$ as
in Fig 1. while those given by VP3 and VP4 as shown in Fig 2.

Here we complete this analysis with some new properties of the
proton-neutron interacting system.
\section{Double beta decay}
\label{sec:level3}
The double beta decay with two anti-neutrinos in the final state
is considered to take place by two consecutive single $\beta^-$
decays. The intermediate state reached after the first beta
decay, consists of an odd-odd nucleus in a pn excited state, one
electron and one anti-neutrino. If in the intermediate state, the
total lepton energy is approximated by the sum of the electron
mass and half of the Q-value of the double beta decay process
($\Delta E$),
the inverse of the process half-life can be factorized as follows
: 
\be
(T^{2\nu}_{1/2})^{-1}=F|M_F|^2,
\ee
where F is a lepton phase integral while the second factor is
determined by the states characterizing the nuclei involved in
the process and has the expression:
\be
M_F=\sum_{k,k'}\frac{{_m}\langle0^+|\beta ^+|0^+_k\rangle_m {_m}
\langle0^+_k|0^+_{k'}\rangle_{d}
{_d}\langle0^+_{k'}|\beta^{\dagger}|0^+\rangle_d}
{E_{k}+\Delta E},
\ee
where the transition operators is defined by eq (2.2).
The initial and final states are the ground states of mother 
($|0^{\dagger}\rangle_m$) and daughter ($|0^{\dagger}\rangle_d$)
nuclei. These states are vacuum states for the pnQRPA phonon
operators built on the top of the corresponding BCS states which
are the static ground states.
The matrix elements describing the first and the second legs of
the process have been calculated in a previous publication
based on the semi-classical approaches (see eqs. 6.25 and 6.32 from
ref.\cite{RaHa}) VP1 and VP2.
In both cases there is only one excited states and therefore the
summation in eq. (3.2) can be omitted. 
The overlap matrix element involved in Eq. (3.2) is function of
the forward (X) and backward (Y) amplitudes of the harmonic boson and has the
expression: 
\be
{_m}\langle0^+_1|0^+_1\rangle_d=X_mX_d -Y_mY_d,
\ee
where the indices m and d specify that the corresponding
states describe the mother and daughter nuclei, respectively.
The energy denominator in Eq (3.2) is:
\be
E_{1}=\omega_m,
\ee
where $\omega_m$, is the energy of the harmonic
mode in the mother nucleus.
The results for the double beta transition amplitude obtained
within the VP1 and VP2 approaches 
are plotted in Fig. 3 a) as function of the proton-neutron pairing
strength k'. 
In the VP1 approach the energy shift $\Delta E$ is taken equal to 1 MeV.
 The same $\Delta E$ is considered for the first part of the k' interval
in Fig. 3 b). For the strength of the proton-neutron pairing $k'\geq1.3MeV$ 
this energy shift is corrected by half the
difference between the static ground state energies in the mother and 
daughter nuclei.
From there we see that the transition amplitude for small
($k'<0.8$MeV) and large ($k'>1.5$MeV) k' is almost constant.
Moreover there is a finite interval for $k'$ where the amplitude
is not defined. The justification for such a window can be
seen in Fig 1. Indeed from the lower panel it is clear
that the VP1 approach yields a transition amplitude which is not
defined beyond $k'=1.05$ since the harmonic mode in the daughter
nucleus collapses.  In the upper panel of Fig.1 it is shown
that the SU(2) mode in the mother nucleus has a non-vanishing energy
only for values of k' larger than 1.2 for $\chi^{\prime}=0.$ and 1.3
for $\chi^{\prime}=0.5$.
\section{Boson Expansion}
\label{sec: level4}
For the VP2 formalism several classical canonical coordinates
have been found. To each set of coordinates  corresponds a specific
quantization scheme. Through the quantization procedure, the 
classical energy function is transformed to an operatorial
function of boson operators. In what follows we shall use two
boson representations of the model Hamiltonian. In each case the
corresponding eigenstates are used to evaluate, by means of eq. (3.2)
, the transition amplitude for the double beta Fermi transition.
\subsection{Holstein Primakoff boson expansion}
The Holstein-Primakoff boson mapping \cite{Hols}
corresponds to the following canonical complex coordinates: 
\be
C^*=\sqrt{2\Omega}V;~~C=\sqrt{2\Omega}V^*,
\ee
Indeed in these coordinates the classical equations of motion
are in the canonical form:
\be
\{C^*,C\}=i,~~
\{C^*,{\cal H}\}=\stackrel{\bullet}{C^*},~~
\{C,{\cal H}\}=\stackrel{\bullet}{C}.
\ee
Through the quantization
\be
(C_1,C_1^*;i\{,\})\rightarrow(B,B^{\dag};[,]),
\ee
the classical SU(2) algebra, generated by the averages of the
fermionic SU(2) algebra on the coherent state $|\Psi_2\rangle$, is
mapped onto a boson SU(2) algebra. Multiplying the two
applications one obtains a mapping of the fermionic algebra onto
a boson SU(2) algebra. The result is conventionally called
as the boson expansion of the initial fermionic generator
operators. Since this boson expansion for the generators of the
SU(2) algebra has been found first by Holstein and  Primakoff we shall
refer to it as to the HP boson expansion. By the mapping
specified by eq.(4.3), any function of the complex coordinates
$C^*, C$ can be transformed into a function of the bosons
$B^\dagger, B$. In particular the classical energy ${\cal H}$, the average
of H (2.1) on $|\Psi_2\rangle$, is transformed in the HP boson
expansion of the  model Hamiltonian:
\bea
H_B^{(HP)}&=&2\epsilon
B^{\dag}B+\lambda_1(B^{\dag}B-\frac{1}{2\Omega} {B^+}^2B^2)
\nonumber\\
&+&\lambda_2[{B^+}^2\sqrt{(1-\frac{\hat{N}+1}{2\Omega})
(1-\frac{\hat{N}}{2\Omega})}+\sqrt{(1-\frac{\hat{N}+1}{2\Omega})
(1-\frac{\hat {N}}{2\Omega})}B^2].
\eea
In the limit of $\Omega$ going to infinity the Hamiltonian $H^{(HP)}_B$ goes
to:  
\be
\widetilde{H}_B^{(HP)}=(2\epsilon +\lambda_1)B^{\dag } B+
\lambda_2({B^{\dag} }^2+B^2).
\ee
We shall refer to $\widetilde{H}_B^{(HP)}$ as the Hamiltonian in the
 zeroth order boson
expansion.
The Hamiltonian in the first order boson expansion is:
\be
H^{(1)}_{HP}=(2\epsilon+\lambda_1)B^{\dag} B
+\lambda_2 \frac{4\Omega-1}{4\Omega}({B^{\dag}}^2+B^2)
-\frac{\lambda_2}{2\Omega}({B^{\dag}}^2
{\hat N}+{\hat N}B^2)-\frac{\lambda_1}{2\Omega}{B^{\dag}}^2B^2.
\ee
The transition operator linking the states described by the
boson Hamiltonians (4.4), (4.5) and (4.6) has the expressions
\bea
\beta^{\dag}&=&\sqrt{2\Omega}\left[U_pV_n\sqrt{1-\frac{B^{\dag} B}{2\Omega}}B
+V_pU_nB^{\dag}\sqrt{1-\frac{B^{\dag} B}{2\Omega}} \right], \\
\beta^{\dag} &=&\sqrt{2\Omega}\left[U_pV_nB
+V_pU_nB^{\dag} \right],\\
\beta^{\dag} &=&\sqrt{2\Omega}\left[U_pV_n(1-\frac{B^{\dag} B}{4\Omega})B
+V_pU_nB^{\dag} (1-\frac{B^{\dag} B}{4\Omega}) \right],
\eea
respectively.
The matrix elements involved in the transition amplitude can be easily
calculated. To save space here we describe  only the
results for the first order expansion. 
Diagonalizing the matrix associated to the 
boson Hamiltonian $H^{(1)}_{HP}$, given in Appendix A, in the
basis 
\be
|m\rangle=\frac{1}{\sqrt{m!}}{B^{\dag} }^m|0\rangle_b,
\ee 
for the case of mother (m) and daughter nuclei (d), one obtains the
eigenstates: 
\be
|0^+_k\rangle_p=\sum_{l}C^{(p)}_{k,l}|l\rangle,~~p=m,d,~~k=0,1,2,..
\ee
Here $|0\rangle_b$ denotes the vacuum state for the boson B. The
corresponding eigenvalues will be denoted by $E^{(p)}_k$.
The label $k=0$ corresponds to the ground state and to simplify
notations we shall omit it. To be more suggestive sometimes the
label "0" is replaced by "g".
The overlap of the states describing the mother and daughter
nuclei have the form:
\be
{_m}\langle 0^+_k|0^+_{k'}\rangle_d=\sum_{l}C^{(m)}_{k,l}C^{(d)}_{k',l}.
\ee
The matrix elements of the transition operator can be easily
calculated once one knows how to calculate the matrix elements
corresponding to the basis states. These are given explicitly in
Appendix A.
The energy denominator in Eq. (2.2), corresponding to the boson
expansion treatment, is defined as follows:
\be
E_{k}=E^{(m)}_k-E^{(m)}_{g}.
\ee
The energy shift $\Delta E$ was taken equal to 1MeV for small values of 
the proton-neutron pairing strength k' while for large k', half
of the difference between the static ground state energies 
of mother and daughter nuclei is added to this value. 
The double beta transition amplitude, calculated by using the eigenstates of
$\widetilde{H}_B^{(HP)}$ is shown in the lower plot of Fig.3,
lowest panel. Since the ground states of the mother and
daughter nuclei contain many boson components the transition
operator, although it is linear in bosons, may excite the ground
states to a many boson state describing the intermediate odd-odd
nucleus.  Therefore for the intermediate states in Eq. (3.2)
 the complete set produced by
the diagonalization procedure was considered. We found
that the first excited state of the above
mentioned complete set contributes the largest part to the
double beta transition amplitude.
The general trends shown in the upper and lower panel are
similar to each other. While the left branches are identical, the
second branch corresponding to the zeroth boson expanded
Hamiltonian is much larger than that produced by the harmonic
approximation defined by using the coherent state for the $SU(2)$
group, as variational state. This indicates that for these values
of k' the anharmonic effects prevail over the harmonic ones.
While in the upper panel we notice an interval where the
transition amplitude is not defined, the
transition amplitude in the lower panel  is vanishing in this interval. 
The reason
is that the mixture of the higher boson states is large and
different components have different phases and therefore their
partial contributions cancel each other. 
The results for the double beta transition amplitude
corresponding to the full HP boson expanded Hamiltonian are
compared in Fig. 4 with those obtained with the first order boson expanded
Hamiltonian . One remarks that sizable differences are noticed
only for large values of k'. It is worth to mention that while
in the lower panel of Fig.3 the
window where the transition amplitude vanishes is for a finite
range of k', here this is reduced just to a point.
As we showed in ref.\cite{RaHa} the HP boson expansion is
defined in that interval of k' where the standard pnQRPA does not
work. However the final boson Hamiltonian can be diagonalized
also for k' belonging to the region where the standard pnQRPA
properly works. In this region however the boson cannot be
interpreted as describing small oscillations around a "deformed"
minimum but around the origin of the phase space. We would like
to mention the fact that the results obtained with the
eigenstates of the full HP Hamiltonian coincide with the exact
result corresponding to the exact eigenstates of the fermionic 
quasiparticle Hamiltonian \cite{RaHa}. 
Comparing the results shown in Figs. 3 (first panel) and 4 one
may say that in the first interval of k' the pnQRPA works
quite well, the corresponding results lie close to the exact
ones. By contrast for large values of k', the results
corresponding to the harmonic approximation are quite different
from the exact ones. The exact result indicate that a big effect
is caused by anharmonicities.
\subsection{Schwinger boson expansion}
In ref. \cite{rad1}we proposed a boson expansion formalism for treating 
the Gamow-Teller double beta transition. The 
underlying idea was that the boson series associated to two
quasiparticle proton neutron operators should be chosen in such
a way that the mutual commutation relations for the fermionic
operators are satisfied in each order of the approximation. This
condition could not be satisfied if we restricted the  space
to the pnQRPA bosons. However, the condition is fulfilled if the
space is extended by including also the quadrupole charge conserving
QRPA bosons. Therefore the boson representation of the
bifermionic proton-neutron operators contains two types of bosons
and therefore we call it Schwinger like boson expansion. 
Schwinger \cite{Schw} was the first who achieved a boson
mapping of the SU(2) algebra  by
using two bosons.
A weak point of boson expansions for many body operators
is that there is no quantitative measure for the "rest" when the
infinite series is truncated. However, in the present case the
evaluation of the "rest" is possible since one knows the exact
result. Therefore it is worth to formulate the boson expansion,
used in ref.\cite{rad1} for GT double beta transition, for the case of
Fermi transition. In this case the charge conserving boson, which
should be added to the proton-neutron QRPA boson in order to
fulfill the condition of preserving the mutual commutation
relations, is the pairing vibrational mode. This mode is a
particle-particle like excitation and its amplitudes are related
to the overlap of the function describing the given nucleus with
a one associated to a system obtained by adding (or
subtracting) a pair of protons or
a pair of neutrons.
It is worth to remark that the daughter nucleus involved in a
Fermi double beta decay is a component of a double phonon
vibrational state obtained by removing a pair of neutrons and
adding a pair of protons. Since the daughter nucleus could be
reached by transforming the mother nucleus in two different
ways, the double beta decay and the double phonon excitation,
one may expect that the anharmonic part of the boson expanded transition
operator for single beta decay leg might excite two boson and
three boson states in the intermediate odd-odd nucleus having
one and two charge conserving phonon factors, respectively.
The pairing vibrational modes are usually defined by the 
proton-proton and neutron-neutron pairing interactions. Therefore
there are proton and neutron vibrational modes which are decoupled
from each other. Within the QRPA approach there are two spurious
states due to the conservation of both the number of protons and the
number of neutrons. When the space of single particle states is
restricted to a single shell there is no physical solution. In
order to allow a non-vanishing energy, one needs at least an additional
level where to promote a pair of particles.
To keep simplicity we maintain the single j picture but
consider the contribution of the proton-neutron particle-hole
and particle-particle interactions to the equations of motion of
the charge conserving operators. In this way, for example, the
single j of the neutron system will play the role of the second
level for the equations of motion of the proton-proton monopole
operators.
Therefore we keep, in the quasiparticle representations of the
model Hamiltonian, the terms with four quasiparticle operators
which contribute to the equations of motion of two quasiparticle
monopole operators:
\bea
H_q &=& 
E^\prime_p \hat{N}_p+E^\prime_n \hat{N}_n
\nonumber \\
&&
-\frac{G_p}{4} 2 \Omega \left[(U_p^4+V_p^4)A^\dagger_{pp}A_{pp}-
U_p^2 V_p^2 (A_{pp}^{\dagger^2}+A_{pp}^2)+4 U_p^2 V_p^2 B_{pp}^\dagger B_{pp}
\right]
\nonumber \\
&&
-\frac{G_n}{4} 2 \Omega \left[(U_n^4+V_n^4)A^\dagger_{nn}A_{nn}-
U_n^2 V_n^2 (A_{nn}^{\dagger^2}+A_{nn}^2)+4 U_n^2 V_n^2 B_{nn}^\dagger B_{nn}
\right]
\nonumber \\
&&
+\lambda_1 A_{pn}^\dagger A_{pn}
+\lambda_2 \left[
A_{pn}^{\dagger^2}+A_{pn}^2-B_{pn}^{\dagger^2} -B_{pn}^2 \right],
\eea
where the following notations have been used:
\bea
E^\prime_p & = & E_p + \frac{G_p}{2} V_p^4 + 2 (U_p^2 - V_p^2) 
(\chi U_n^2 - \chi_1 V_n^2),
\\
E^\prime_n & = & E_n + \frac{G_n}{2} V_n^4 - 2 V_p^2 (U_n^2 - V_n^2)
(\chi + \chi_1),
\\
E_p & = & \frac{G_p \Omega}{2},~~
E_n  =  \frac{G_n \Omega}{2},
\\
\Delta_p & = & G_p \Omega U_p V_p,~~
\Delta_n  =  G_n \Omega U_n V_n,
\\
V_n & = & \sqrt{\frac{N_n}{2 \Omega}},~~
V_p  = \sqrt{\frac{N_p}{2 \Omega}},
\\
A^{\dagger}_{\tau \tau}&=&\frac{1}{{\hat
j}}\sum_{m}a^{\dagger}_{\tau m}a^{\dagger}_{\widetilde{\tau m}}
,~~\tau=p,n.
\eea
Assuming quasiboson commutation relations for the operators 
$A^{\dagger}_{\tau \tau}, A^{\dagger}_{\tau \tau}$ one obtains
the following linearized equations of motion:
\bea
\left[ H_q, A_{pp}^\dagger \right] & = &
\left(2 (E^\prime_p - E_p)+\frac{\Delta_p^2}{E_p}\right) A^\dagger_{pp}
-(\chi + \chi_1) \frac{\Delta_p \Delta_n}{E_p E_n} A^\dagger_{nn}
\nonumber \\
&&
+\frac{\Delta_p^2}{E_p} A_{pp}
-(\chi + \chi_1) \frac{\Delta_p \Delta_n}{E_p E_n} A_{nn},
\\
\left[ H_q, A_{nn}^\dagger \right] & = &
-(\chi + \chi_1) \frac{\Delta_p \Delta_n}{E_p E_n} A_{pp}^\dagger 
+\left(2 (E^\prime_n - E_n)+\frac{\Delta_n^2}{E_n}\right)A^\dagger_{nn}
\nonumber \\
&&
-(\chi + \chi_1) \frac{\Delta_p \Delta_n}{E_p E_n} A_{pp}
+\frac{\Delta_n^2}{E_n}A_{nn},
\\
\left[ H_q, A_{pp} \right] & = &
-\frac{\Delta_p^2}{E_p} A^\dagger_{pp}
+(\chi + \chi_1) \frac{\Delta_p \Delta_n}{E_p E_n}A^\dagger_{nn}
\nonumber \\
&&
-\left(2 (E^\prime_p - E_p)+\frac{\Delta_p^2}{E_p}\right) A_{pp}
+(\chi + \chi_1) \frac{\Delta_p \Delta_n}{E_p E_n}A_{nn},
\\
\left[ H_q, A_{nn} \right] & = &
(\chi + \chi_1) \frac{\Delta_p \Delta_n}{E_p E_n}A^\dagger_{pp}
-\frac{\Delta_n^2}{E_n}A^\dagger_{nn}
\nonumber \\
&&
+(\chi + \chi_1) \frac{\Delta_p \Delta_n}{E_p E_n}A_{pp}
-\left(2 (E^\prime_n - E_n)+\frac{\Delta_n^2}{E_n}\right)A_{nn}.
\eea
These equations allows us to determine the  operator
\be
\Gamma^{\dagger}_{\tau
\tau}=X_pA^{\dagger}_{pp}+X_nA^{\dagger}_{nn}
-Y_pA_{pp} -Y_nA_{nn}.
\ee
which fulfills the restrictions
\bea
\left[H_q,\Gamma^{\dagger}_{\tau\tau}\right]&=&\omega 
\Gamma^{\dagger}_{\tau\tau},
\nonumber \\
\left[\Gamma_{\tau\tau},\Gamma^{\dagger}_{\tau\tau}\right]&=&1.
\eea
The first equation (4.26) provides an homogeneous system of linear
equations for the amplitudes X and Y. To be able to solve these
equations, the determinant of the coefficients has to be zero. This determines
the excitation energy $\omega$:
\bea
&&\left[
4(E^\prime_p - E_p)^2 - \omega^2 + 4 \frac{\Delta_p^2}{E_p} (E^\prime_p - E_p)
\right] 
\left[
4(E^\prime_n - E_n)^2 - \omega^2 + 4 \frac{\Delta_n^2}{E_n} (E^\prime_n - E_n)
\right] \nonumber \\
& & - 16(E^\prime_p - E_p) 
 (E^\prime_n - E_n)
(\chi + \chi_1)^2 \frac{\Delta_p^2 \Delta_n^2}{E^2_p E^2_n}=0.
\eea
The amplitudes are obtained up to a multiplicative constant which is fixed
by the normalization restriction given by the second eq. (4.26 ),
which reads:
\be
2(X^2_p+X^2_n-Y^2_p-Y^2_n)=1.
\ee
In a similar way one determines the QRPA equations for the
amplitudes and energy of the proton-neutron phonon operator:
\be
\Gamma^{\dag}_{pn}=X_{pn}A^{\dag}_{pn}-Y_{pn}A_{pn}.
\ee
The energy has the expression
\be
\omega=\left[(E^{\prime}_p+E^{\prime}_n+\lambda_1)^2-
4\lambda_2^2\right]^{\frac{1}{2}}
\ee
The corresponding phonon amplitudes are:
\bea
X_{pn}&=&\left[1-\frac{E^{\prime}_p+E^{\prime}_n+\lambda_1-\omega}{4\lambda_2^2}
\right]^{-\frac{1}{2}},\nonumber \\
Y_{pn}&=&-\frac{E^{\prime}_p+E^{\prime}_n+\lambda_1-\omega}{2\lambda_2}X_{pn}.
\eea
Following the procedure described in detail in ref.\cite{rad1} one
obtains the following boson expansion of Schwinger type for the
operators involved in the transition operator:
\bea
B^{\dag}_{pn}&=&b\Gamma^{\dag}_{pn}\Gamma^{\dag}_{\tau\tau}+c
\Gamma^{\dag}_{pn}\Gamma_{\tau\tau}+
d\Gamma^{\dag}_{\tau\tau}\Gamma_{pn}+
e\Gamma_{pn}\Gamma_{\tau\tau},\nonumber \\
A^{\dag}_{pn}&=&a_{10}\Gamma^{\dag}_{pn}+a_{01}\Gamma_{pn}+
a_{30}\Gamma^{\dag}_{pn}
\Gamma^{\dag}_{\tau\tau}\Gamma^{\dag}_{\tau\tau}
+a_{21}\Gamma^{\dag}_{pn}
\Gamma^{\dag}_{\tau\tau}\Gamma_{\tau\tau}+a_{12}
\Gamma^{\dag}_{pn}
\Gamma_{\tau\tau}\Gamma_{\tau\tau}
\nonumber \\
&+&a_{\bar{21}}\Gamma^{\dag}_{\tau\tau}
\Gamma^{\dag}_{\tau\tau}\Gamma_{pn}+
a_{\bar{12}}
\Gamma^{\dag}_{\tau\tau}
\Gamma_{pn}\Gamma_{\tau\tau}
+a_{03}\Gamma_{pn}
\Gamma_{\tau\tau}\Gamma_{\tau\tau}.
\eea
The expansion coefficients are given in Appendix B.
These boson expansion induces anharmonic components for the
transition operator which excite the ground state f the mother
nucleus to some many boson states. Note that within the standard
pnQRPA approach, only the one phonon state is accepted as
intermediate state, all other states produce vanishing matrix
elements. We denote by $M^{(k)}_F$, k=1,2,3 the transition
amplitudes for the double beta Fermi transition determined by
one, two and three boson states of the intermediate nucleus.
Within the first order boson expansion the result for the total amplitude
is: 
\be
M_F=M^{(1)}_F+M^{(2)}_F+M^{(3)}_F,
\ee
where the partial amplitudes are given explicitly in Appendix C.

Within the standard pnQRPA formalism, the double beta transitions
to excited states of the daughter nuclei are forbidden. If the
first order boson expansion of the transition operator is
considered then the transition to the first and double phonon
states are allowed. Here we study the transition to the double
phonon pairing vibration. The amplitude for this transition has
the expression:
\bea
M_F^{2ph} & = & 
  2\sqrt{2}\Omega \frac{(U_n^m V_p^m a_{01}^m + 
          U_p^m V_n^m a_{10}^m)O^{(1)}_{md}  (U_n^d V_p^d a_{12}^d + 
            U_p^d V_n^d a_{\bar{21}}^d)}
{(\omega_{pn}^m + 
              \Delta E)^3 }
\nonumber \\
&&+2\sqrt{2}\Omega \frac{(U_n^m V_p^m a_{03}^m + U_p^m V_n^m a_{30}^m) 
        O^{(3)}_{md}
        (U_n^d V_p^d a_{10}^d + 
      U_p^d V_n^d a_{01}^d)}
{(\omega_{pn}^m + 2 \omega^m + \Delta E)^3}.
\eea
The overlap matrices $O^{(1)}$ and $O^{(3)}$ are defined in Appendix
C. The superscripts m and d suggest that the given quantities
characterize the mother and daughter nuclei, respectively.

The results corresponding to transition operators defined in the
standard pnQRPA (i.e. linear in the pn bosons), and higher order
pnQRPA by including quadratic and cubic terms in the bosons expansion 
are shown in
Fig.5 together with the exact result.
Of course since the pn boson collapses at about $k'=1.05MeV$
this boson expansion is justified only for the interval
below this value. One notes that the quadratic terms in
bosons included in the transition operator modifies the pnQRPA
result for the double beta transition amplitude such that the
final result is in excellent agreement with the exact one. Also
one notices that the third order boson terms are practically not
modifying the result obtained with the linear and quadratic
terms. One may conclude that at least for the single j shell
considered here and the double beta Fermi transition the first order
boson expansion is approaching very well the exact result.
Of course important deviations appear when we approach the 
critical value of k' where the pn boson collapses.
For example  for k' where the pnQRPA amplitude is equal to zero
the exact result is about 0.9 MeV$^{-1}$.
As we have already mentioned within the pnQRPA approach the
transition to an excited state of the daughter nucleus
is forbidden. However this is allowed within the boson expansion
formalism. Here we calculated the amplitude for the transition
$0^+_i\rightarrow(2ph)_f$ from the ground state of the mother to
the two vibrational phonon states of the daughter. Such a state
may be alternatively reached in a process which removes a pair
of neutrons and adds a pair of protons to the mother nucleus. 
Therefore if pairing vibrational phonons play a role by
inducing anharmonic effects in the $\beta^-$ transition operator the
double phonon state might be populated in a double beta decay.
The results are shown in Fig 6. From there one remarks that the
process amplitude is reasonable large and if that order of
magnitude  persists for realistic calculations one might hope
that such transitions can be experimentally identified.
Coming back to Fig 5, one notices that around $k'=1.2$ the exact
amplitude suggests a phase transition since its first derivative
has a discontinuity there. In the RPA procedure such a
transition is associated to a Goldstone mode which appears
whenever a new symmetry open up. What happens however in the
case of the exact solution? To give an answer to this
question we plotted in Fig 7 the first five eigenvalues of the
HP boson expanded Hamiltonian for the mother (upper panel) and
the daughter (lower panel) nuclei. Also in Fig. 8 we give the first two
leading amplitudes of the first four eigenstates.
In Fig. 7 one sees that not far from the turning point 
in Fig. 5 the first excited state is almost degenerate with the
ground state. The virtual transition to such a state is almost
forbidden, despite the fact that its
excitation energy is so small, since the matrix elements of the
transition operator are suppressed.
This happens due to the fact that two amplitudes of the
wave functions (see Fig. 8) in
this region are becoming equal in magnitude but of opposite phase.
Indeed for a small k' the dominant component of the ground state is
the vacuum state while for k' greater than 1.75 the third
component, corresponding to the two phonon states, becomes
dominant. That is in fact a signature for a super-deformed ground
state. A similar picture holds also for the first excited state.
For small k' this is a one boson state while for k' larger than
1.5 it is a three boson state. It is interesting to see that for
the mother nucleus the initial components ordering of the third
excited state is recovered by increasing k' beyond the value
2.4. It seems that beyond the k'=1.2 the contributions of the
higher boson states to the total transition amplitude are
constructively interfering.

Let us now conclude the present Section. While in the preceding
section we treated the double beta transition amplitude within
an QRPA like approaches provided by VP1 and VP2, here boson
expansion procedures have been used in order to incorporate
higher RPA correlations. In the region of large
particle-particle interaction strength k' we have used a full 
HP boson expanded Hamiltonian
and two truncated HP boson expansions. In the interval of small
strengths for the particle-particle interaction  k' a Schwinger
like boson expansion have been used. In both cases we
compared the results of truncated boson expansions with the
corresponding RPA like approach as well as with the exact
results. In both cases the first order boson expansions for the
transition operator provides a description which is very close
to the exact picture.
While in the interval of small particle-particle interaction strength k'
the VP1 result is not bad comparing it to the exact one, in the
complementary region the result produced by the VP2 exhibits large deviation
from the exact result. This reflects the fact that the exact
result includes large contributions coming from anharmonicities.
The transition from one regime to another could be interpreted
by inspecting the structure of the first exact eigenstates of the
model Hamiltonian. Indeed going beyond $k'\approx1.5$ the
dominant component in the ground state is the two phonon state
while in the first excited state the three phonon state prevails.
\section{New features of the RPA vacua and static ground states}
\label{sec: level5}
The BCS approximation determines variationally the ground state
of a system of nucleons interacting among themselves through
pairing force. This approximation determines simultaneously the
optimal ground state and a new type of particle excitations which
admit the found ground state as vacuum . In a time
dependent variational formalism the BCS ground state turns out to be
the static ground state with respect to which one could define the
small oscillations which account for the 
quasiparticle two body interaction. This approach is fully equivalent to the
QRPA formalism. The QRPA formalism finds also an excitation
operator which involves excitations of many quasiparticles from
the ground state. However the vacuum state, or in other words the
new ground state involving correlations corresponding to these
degrees of freedom, is not a solution of a variational equation.

In ref.\cite{RaHa}, using 4 different trial functions we determined
variationally four static ground states and moreover four
RPA like phonon states. The vacuum states for the phonon
operators yielded by VP1 and VP2 can be analytically obtained.
Indeed the equation expressing the condition of being vacuum
for  the phonon operators can be solved and the results are:
\bea
|0\rangle_w&=&{\cal N}_W\exp\left[{\frac{Y_w}{2X_w}(A^{\dag}_{pn})^2}\right],
 \\
|0\rangle_{su(2)}&=&{\cal N}_{su(2)}\exp\left[{\frac{Y_{su(2)}}
{2X_{su(2)}}(A^{\dag}_{pn})^2}\right].
\eea
The indices w and su(2) suggest that the amplitudes X and Y
characterize the phonon operator describing small oscillations
around a static ground state represented by a coherent state of
the Weyl and SU(2) groups, respectively. These amplitudes were
determined in our previous publication (see eq. 6.20 and 6.28 of
ref.\cite{RaHa}).
Assuming for the operators $A^{\dag}$ involved in (5.1,5.2) 
quasiboson commutation relations one obtains the following
expressions for the norms.   
\bea
{{\cal N}_w}^{-2}&=&\sum_{l=0}^{\Omega}(\frac{Y_w}{2X_w})^{2l}
\frac{2l!}{(l!)^2},
\nonumber\\
{{\cal N}_{su(2)}}^{-2}&=&\sum_{l=0}^{\Omega}(\frac{Y_{su(2)}}{2X_{su(2)}})^{2l}
\frac{2l!}{(l!)^2},
\eea
It is worth noting that both vacua are exponential functions of
$(A^{\dag})^2$ . On the other hand so are the static ground
states yielded by the VP3 and VP4 approaches:
\bea
|\Psi_3\rangle&=&{\cal N}_{su(1,1)}
\exp{\stackrel{\circ}{\alpha}(A^{\dag}_{pn})^2},
\alpha=\frac{e^{i\phi}}{2}tanh(2\rho) \\
|\Psi_4\rangle&=&{\cal N}_4\exp{z(A^{\dag}_{pn})^2},~ z=\rho e^{i\phi}.
\eea
The norm ${\cal N}_{su(1,1)}$ has been analytically calculated
\be
{\cal N}_{su(1,1)}=\frac{1}{\sqrt{cosh(2\rho)}},
\ee
while ${\cal N}_4$  has only been numerically determined.
The parameters entering the defining equations (5.4) are those
which produce a minimum energy for our system.
The question we address in this paper is how one compares the
vacua of the phonons defined by means of the VP1 and VP2
formalisms to the static ground states of VP3 and VP4.
If they are close to each other one could state that the
solutions of the variational equations VP3 and VP4, i.e. the
static ground states, approximate the vacua of VP1 and VP2 which
are the dynamic ground state of the system.
For an easier presentation it is convenient to use the unified
notation
\be
|\Phi_k\rangle=
{\cal N}_k\exp\left[{\stackrel{\circ}{\alpha}_k(A^{\dag}_{pn})^2}
\right ]~~ ,k=1,..,4.
\ee
The indices $k=1,2$ correspond to the vacua defined within the
VP1 and VP2 formalisms, respectively, while $k=3,4$ give the static
ground states provided by VP3 and VP4, respectively.
In Figs 9,10 and 11 we compare the norms,
exponents and the products ${\cal N}\alpha $ of the four
functions. Actually the first and last quantities are just the
weights of
quasiparticle vacuum and double phonon excitations, respectively, for the
four states mentioned above. From these figures one notes that
for situations when the first two components of these four
functions are dominant one may state the following. For
$k'\leq0.8MeV$ the vacuum state defined by the VP1 approach is
quite well approximated by the static ground state corresponding
to VP3. The static ground state of the VP4 approach
approximates very well the vacuum state of VP1 and reasonable well
the one corresponding to VP2 in the interval of k' from 1.3 to
1.8 MeV. This result is very important since one could think of
formulating a theory which improves the RPA approach by that the
vacuum of the phonon operator satisfy the equations provided by
a variational principle.
\section{Conclusions}
\label{sec: level6}
The main results obtained in the previous sections can be
summarized as follows.
The model Hamiltonian for a single j shell was treated within a
harmonic picture for values of the particle-particle interaction
strength k' below and beyond the critical value, where the standard
pnQRPA ceases to be any longer valid. The energies and the
states describing the proton and neutron systems are obtained
through a time dependent variational principle choosing as trial
function coherent states of the symmetry groups of Weyl and
$SU(2)$, respectively. The results are used to calculate the
transition amplitude for double beta Fermi transitions. In the
range for larger k' values one may define a Holstein-Primakoff
boson representation of the model Hamiltonian which is further
on abusively used also for the complementary interval. The transition
amplitude for the double beta decay was calculated using the
zeroth, the first order as well as the full HP boson
expansion for the model Hamiltonian. The last procedure
reproduces the exact result.
For the first part of interval we also considered a Schwinger type
boson expansion which is similar to the one  proposed for the 
treatment of the Gamow Teller double
beta decay, by two of the present
authors (A. A. R. and A. F.). By the boson expansion analysis we concluded that
the first order truncation provides a good approximation to the
exact results. In the case of the Schwinger boson expansion we
calculated the transition to the double phonon vibrational
state. Since the daughter nucleus
can be reached by a double phonon excitation of the mother
nucleus, the vibrational states should play an important role in
explaining quantitatively the decay rates.
The two harmonic approximations, for small and large values of 
the particle-particle two body interaction strength k', define phonon
operators whose vacua are dynamical ground states. These states
can be analytically expressed and they exhibit a striking
resemblance with the
variational states used by the VP3 and VP4 approaches. The question is how
one compares these states with each other. We identified
intervals where the dynamic ground states are approximated quite
well by the states determined variationally. This is in fact the
first step toward solving an old standing problem of determining
variationally the RPA ground state.
We hope that the results of this paper will stimulate further
studies both for the field of double beta decay in the region of
large particle-particle interaction but also for 
the static and dynamic ground states of many body systems.
\section{Appendix A}
\label{sec: levelA}
The matrix elements of the first order expanded Hamiltonian in
the many boson basis have the following explicit expressions:
\bea
\langle m|H^{(1)}_{HP}|m\rangle &=& [2\epsilon+\lambda_1)m-
\frac{\lambda_1}{2\Omega}m(m-1),\nonumber \\
\langle m|H^{(1)}_{HP}|m+2\rangle &=&\lambda_2\frac{4\Omega-1}{4\Omega}
\sqrt{(m+1)(m+2)}-\frac{\lambda_2}{2\Omega}m\sqrt{m+1)(m+2)},\
\nonumber \\
 \langle m+2|H^{(1)}_{HP}|m\rangle &=&\langle
m|H^{(1)}_{HP}|m+2\rangle .
\eea
The matrix elements for the first leg and second leg of the
double beta transition are given by:
\bea
{_m}\langle 0^{\dag}|\beta^{\dag}|0^{\dag}_k\rangle _m &=&
\sqrt{2\Omega}\sum_{l,l'}C^{(m)}_{0,l}C^{(m)}_{kl'}
\nonumber \\
&\times&\left[U_pV_n(1-\frac{m}{4\Omega})
\sqrt{l+1}\delta_{l',l+1}
+V_pU_n(1-\frac{m-1}{4\Omega})\sqrt{l}\delta_{l',l-1}\right],\nonumber
\\
{_d}\langle 0^{\dag}_k|\beta^{\dag}|0^{\dag}\rangle _d&=&
\sqrt{2\Omega}\sum_{l,l'}C^{(d)}_{0,l'}C^{(d)}_{kl}
\nonumber \\
&\times&\left[U_pV_n(1-\frac{m}{4\Omega})
\sqrt{l+1}\delta_{l',l+1}
+V_pU_n(1-\frac{m-1}{4\Omega})\sqrt{l}\delta_{l',l-1}\right] .
\eea
\section{Appendix B}
\label{sec: levelB}
Here we list the coefficients defining the boson expansions (4.32).
\bea
b&=&\frac{2}{{\hat j}}(X_pY_{pn}+Y_nX_{pn}),\nonumber \\
c&=&\frac{2}{{\hat j}}(X_nX_{pn}+Y_pY_{pn}), \nonumber \\
b&=&\frac{2}{{\hat j}}(X_pX_{pn}+Y_nY_{pn}), \nonumber \\
b&=&\frac{2}{{\hat j}}(X_nY_{pn}+Y_pX_{pn}), \nonumber \\
a_{10}&=&X_{pn},
\nonumber \\
a_{01}&=&Y_{pn},
\nonumber \\
a_{30}&=&-\frac{2X_{pn}}{\Omega}(X_pY_p+X_nY_n),
\nonumber \\
a_{21}&=&-\frac{2X_{pn}}{\Omega}(X_p^2+X_n^2+Y_p^2+Y_n^2),
\nonumber \\
a_{12}&=&-\frac{2X_{pn}}{\Omega}(X_pY_p+X_nY_n),
\nonumber \\
a_{\bar{21}}&=&-\frac{2Y_{pn}}{\Omega}(X_pY_p+X_nY_n),
\nonumber \\
a_{\bar{12}}&=&-\frac{2Y_{pn}}{\Omega}(X_p^2+X_n^2+Y_p^2+Y_n^2),
\nonumber \\
a_{03}&=&-\frac{2Y_{pn}}{\Omega}(X_pY_p+X_nY_n).
\eea
In what follows we shall briefly describe the procedure
presented in ref.\cite{rad1,rad2}. From the eqs.. (4.32) one derives the equations:
\bea
b&=&\langle0|\left[\left[B^{\dag}_{pn},\Gamma_{\tau\tau}\right],\Gamma\right]
|0\rangle,\nonumber\\
a_{30}&=&\frac{1}{2}\langle0|\left[\Gamma_{\tau\tau},\left[\Gamma_{\tau\tau},
\left[\Gamma_{pn},A^{\dag}_{pn}\right]\right]\right]|0\rangle.
\eea
In the above equations, all commutators are exactly evaluated
except for the last one for which the quasi-boson approximation
is used.
It is worth noting that the coefficients calculated in this way
determine a boson representation for the bi-fermionic operators
which satisfy the mutual commutation relations in the first order.
\section{Appendix C}
\label{sec: levelC}
Here we list the transition amplitudes corresponding to the
pnQRPA and first order boson expansion:
\bea
M_F^{1} & = & 
2 \Omega \frac{(U_n^m V_p^m Y_{pn}^m + U_p^m V_n^m
X_{pn}^m)O^{(1)}_{md} (U^d_nV^d_pX_{pn}^d+
          U_p^d V_n^d Y_{pn}^d)}
{\omega_{pn}^m + \Delta E},
\nonumber \\
M_F^{(2)} & = &
  2 \Omega \frac{(-V_n^m V_p^m e^m + 
        U_p^m U_n^m b^m)O^{(2)}_{md}(-V_n^d V_p^d b^d + 
          U_p^d U_n^d e^d) }
{\omega_{pn}^m + \omega^m + \Delta E},
\nonumber \\
M_F^{(3)} & = &
  4 \Omega \frac{(U_n^m V_p^m a^m_{03} + 
        U_p^m V_n^m a^m_{30})O^{(3)}_{md}(U_n^d V_p^d a^d_{30} + 
          U_p^d V_n^d a^d_{03}) }
{\omega_{pn}^m + 2 \omega^m + \Delta E}.
\eea
The overlap matrices have the expressions:
\bea
O^{(1)}_{md}&=&(X_{pn}^m X_{pn}^d - 
        Y_{pn}^m Y_{pn}^d),
\nonumber \\
O^{(2)}_{md}&=&X_p^m X_p^d + 
        X_n^m X_n^d - 
        Y_p^m Y_p^d - 
        Y_n^m Y_n^d,
\nonumber \\
O^{(3)}_{md}&=&(X_{pn}^m X_{pn}^d - 
        Y_{pn}^m Y_{pn}^d)
        (X_p^m X_p^d + X_n^m X_n^d - 
        Y_p^m Y_p^d - Y_n^m Y_n^d)^2.
\eea

\begin{figure}[ht!]
\vskip 19cm 
\includegraphics{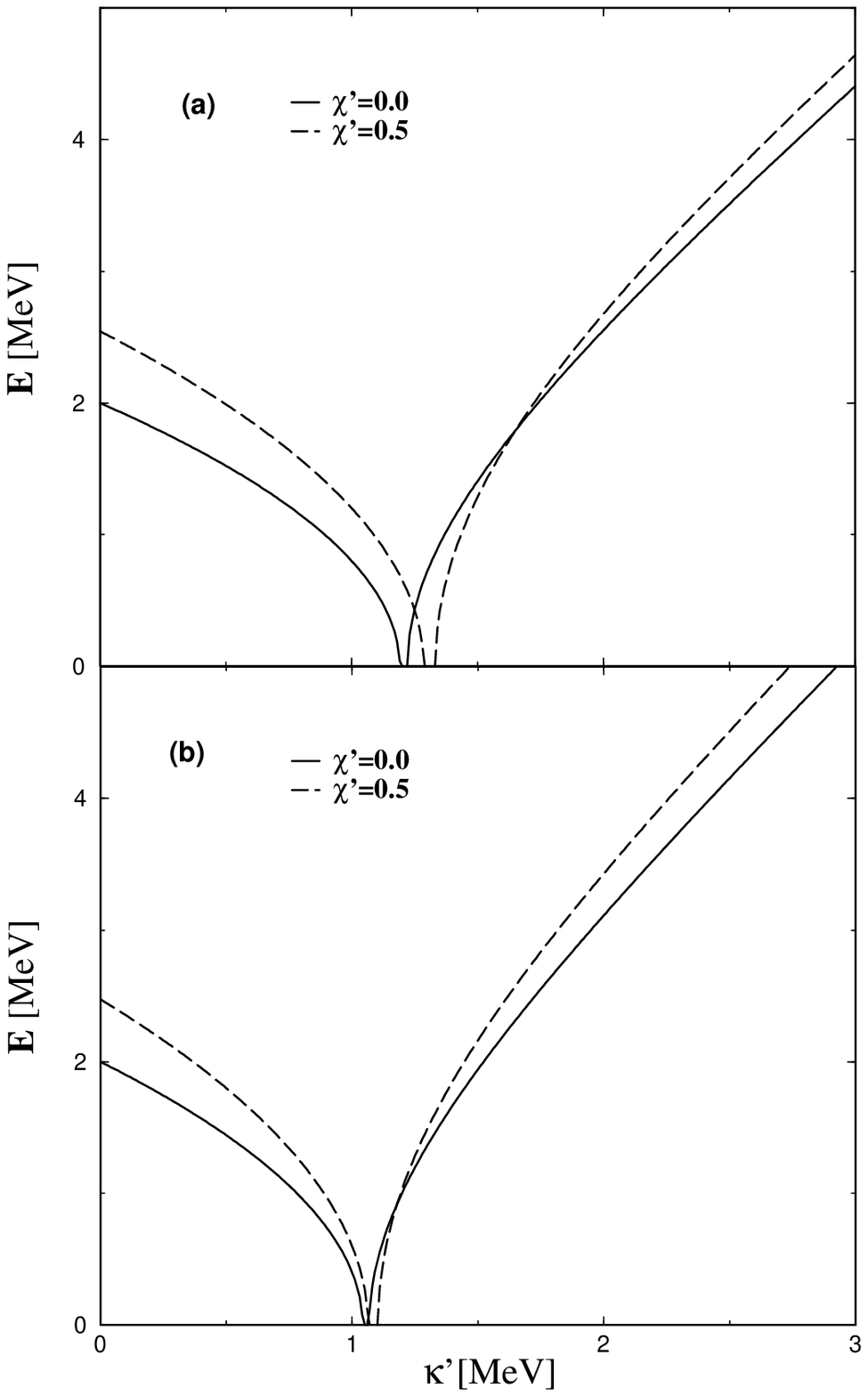}
\caption{The energies of the harmonic modes corresponding to the VP1 
(left branch) and VP2 (right branch) defined in eqs. (2.7) and (2.8) 
are plotted as functions of the proton-neutron pairing strength (eqs.
(2.2) and (2.11)) k' 
for  the mother (a) and the daughter (b) nuclei. Energies as well as 
strength $\chi'$ are given in units of MeV}
\end{figure}
\clearpage

FIGURE 2
\begin{figure}[ht!]
\vskip 20.2cm 
\includegraphics{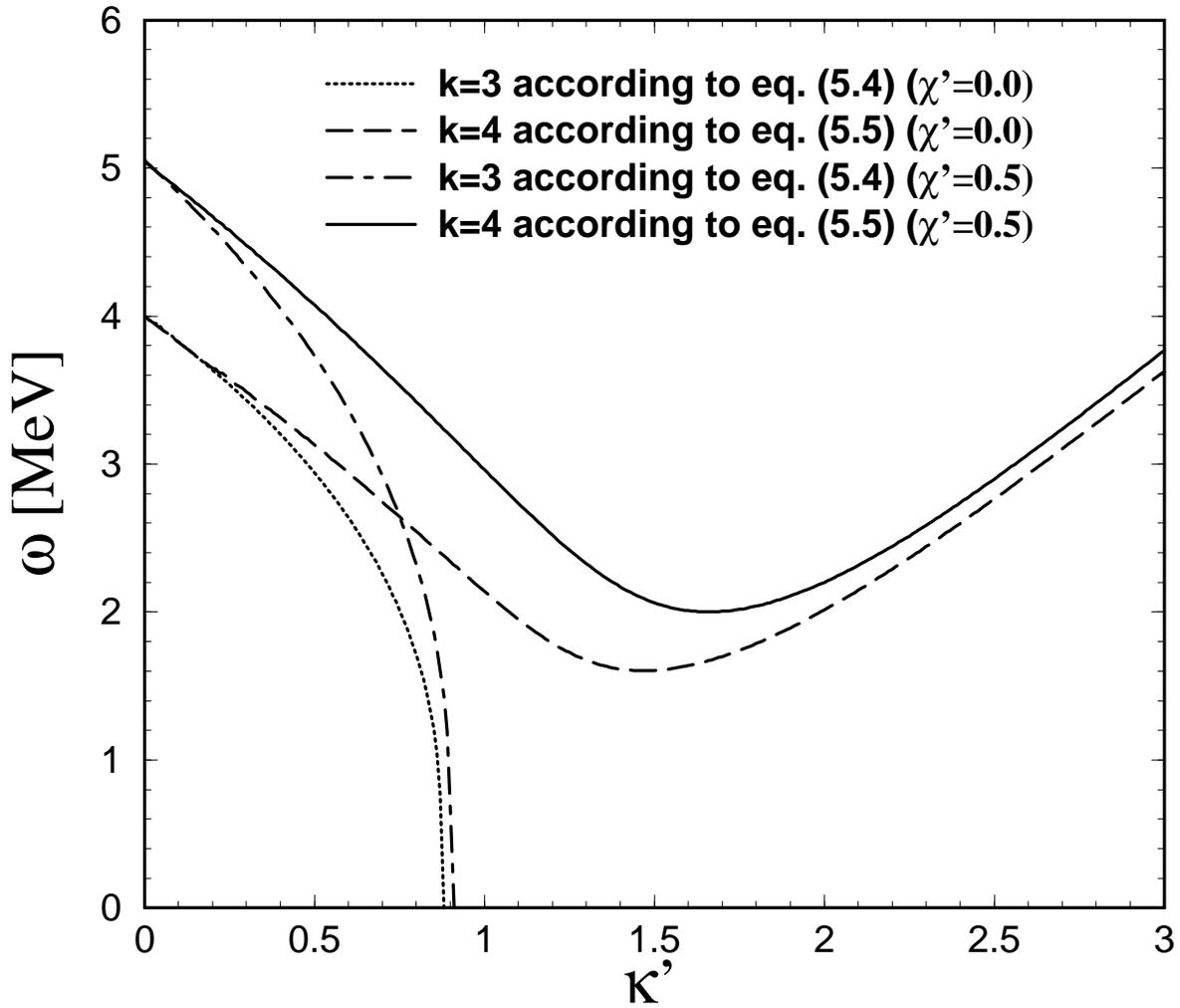}
\caption{Energies of the harmonic modes corresponding to VP3 
(point and dashed-point lines) and VP4 (full and dashed lines) are given 
in units of MeV as functions of proton-neutron pairing strength k'}
\end{figure}
\clearpage

FIGURE 3
\begin{figure}[ht!]
\vskip 19cm 
\includegraphics{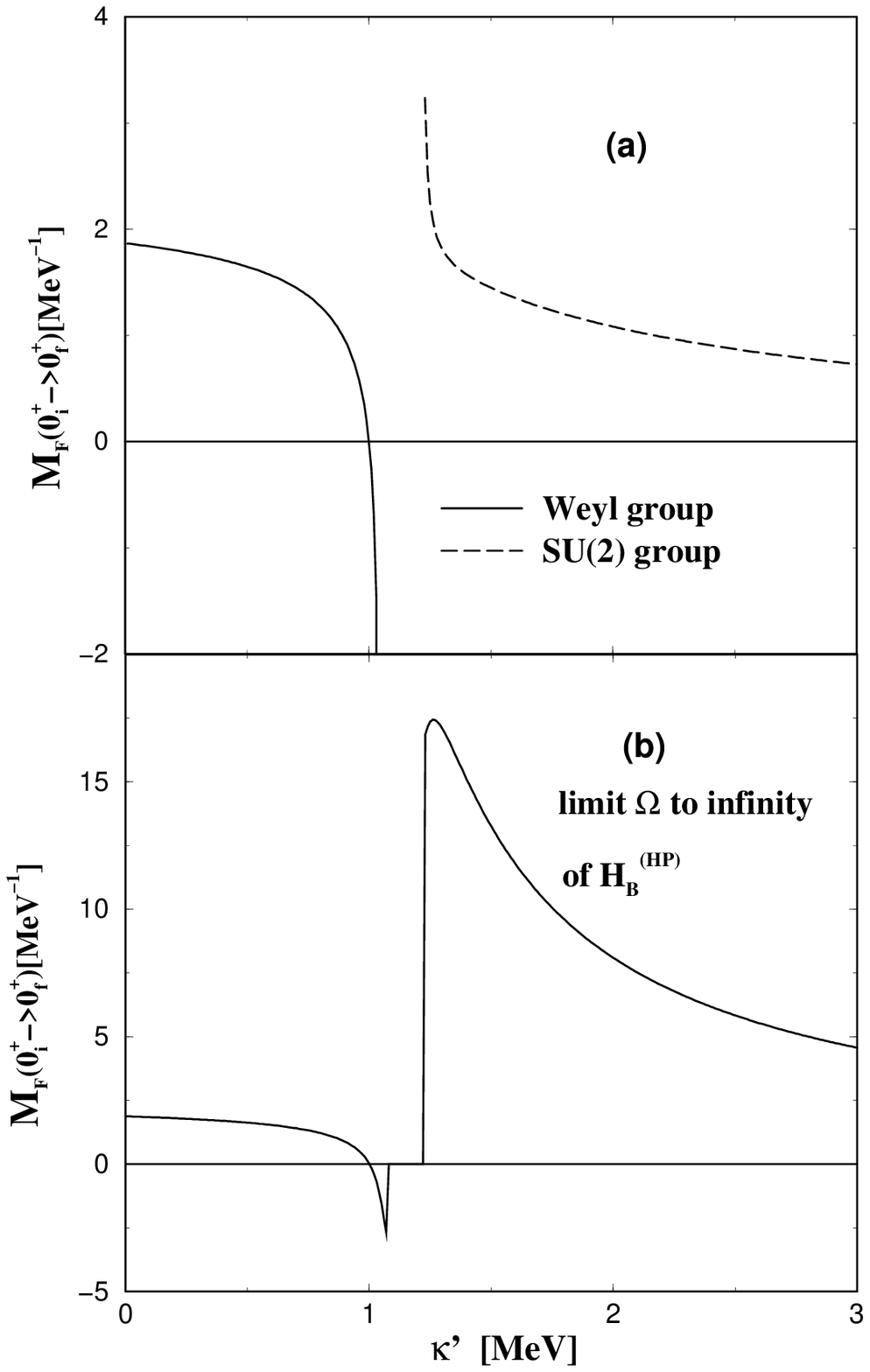}
\caption{The double beta Fermi transition amplitudes obtained with the 
initial, intermediate and final states provided by VP1 (full line)  
VP2 (dashed line) formalism (a) 
and by diagonalizing the limit $\Omega $ goes to infinity of the HP boson 
expanded Hamiltonian (b) are given as function of the particle-particle 
interaction strength k'.}
\end{figure}
\clearpage

FIGURE 4
\begin{figure}[ht!]
\vskip 19cm 
\includegraphics{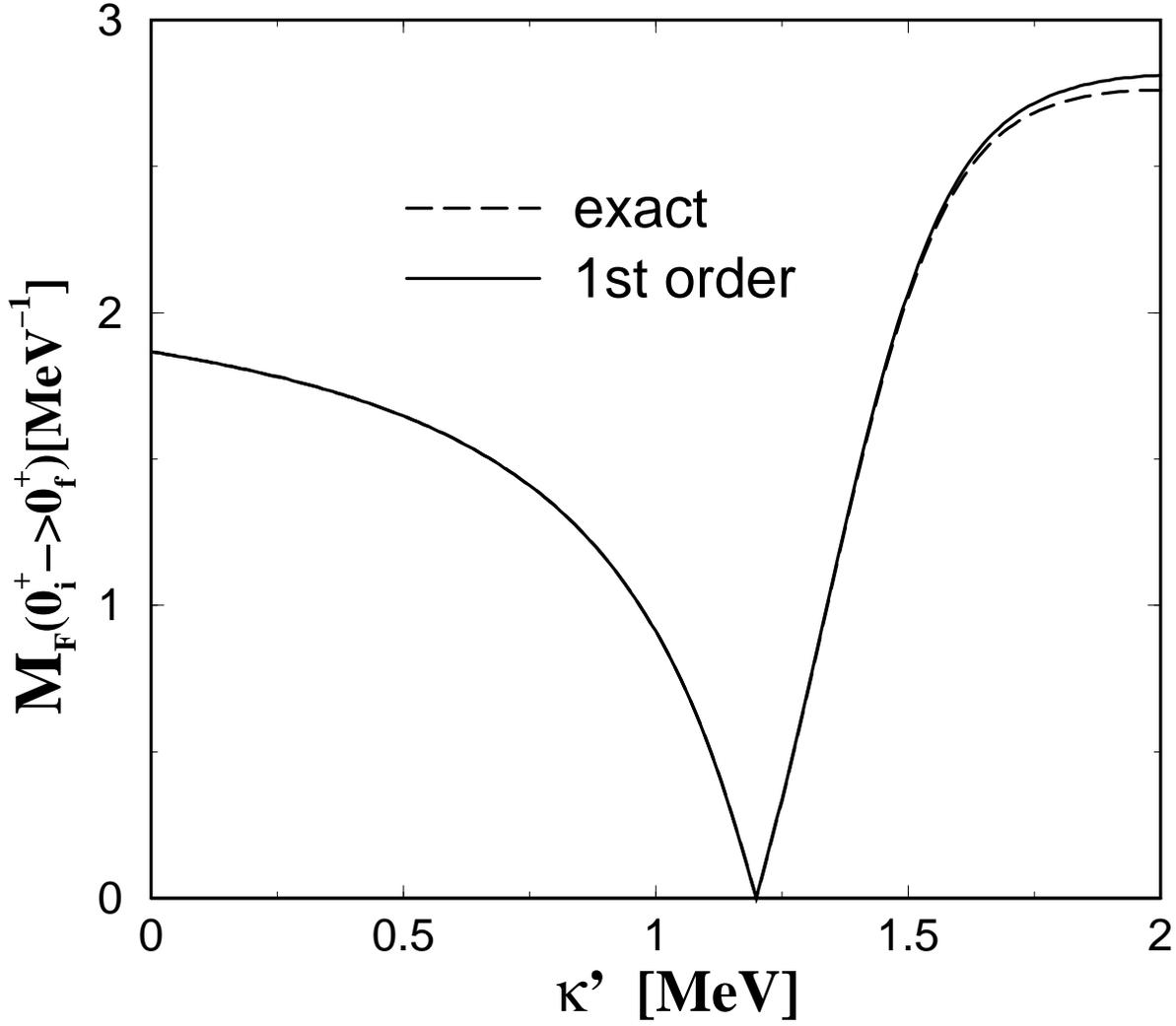}
\caption{The double beta transition amplitude obtained by considering the
states for mother, daughter and intermediate odd-odd nuclei as eigenstates 
of the Holstein-Primakoff boson expanded Hamiltonian (full line) and of 
the first order 
truncated HP Hamiltonian (dashed line) as  function of the
proton-neutron pairing strength k' defined as in eq. (2.11)}
\end{figure}
\clearpage

FIGURE 5
\begin{figure}[ht!]
\vskip 17.2cm 
\includegraphics{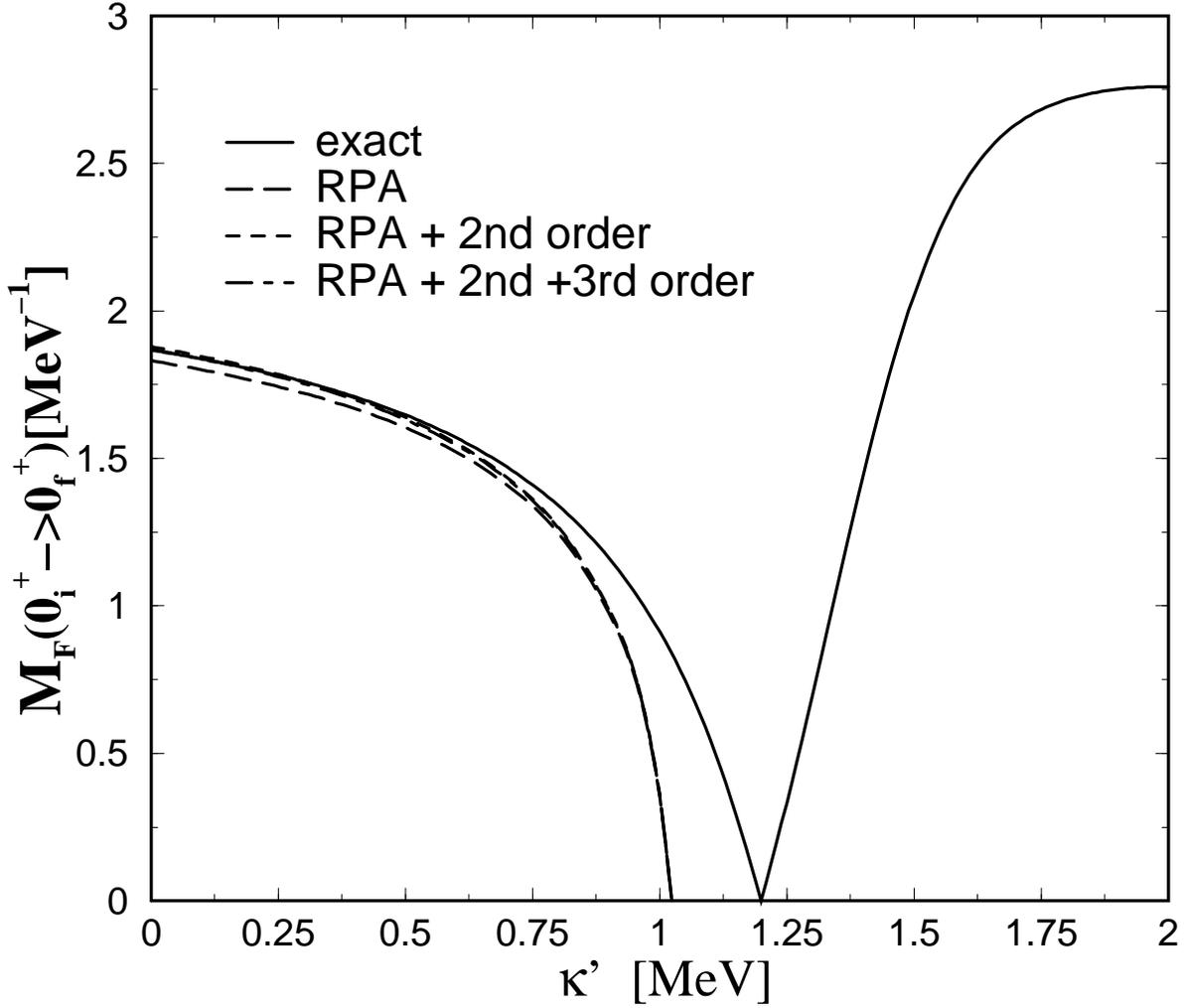}
\caption{The transition amplitude for double beta Fermi decay calculated
 with the standard pnQRPA (long dashed line), pnQRPA plus the second order
 corrections(dashed line) and pnQRPA plus second and third order corrections
 (dashed-point-point line)in the transition operator is represented as 
function of k'. For comparison the exact result is also presented (full line).}
\end{figure}
\clearpage

FIGURE 6
\begin{figure}[ht!]
\vskip 20.2cm 
\includegraphics{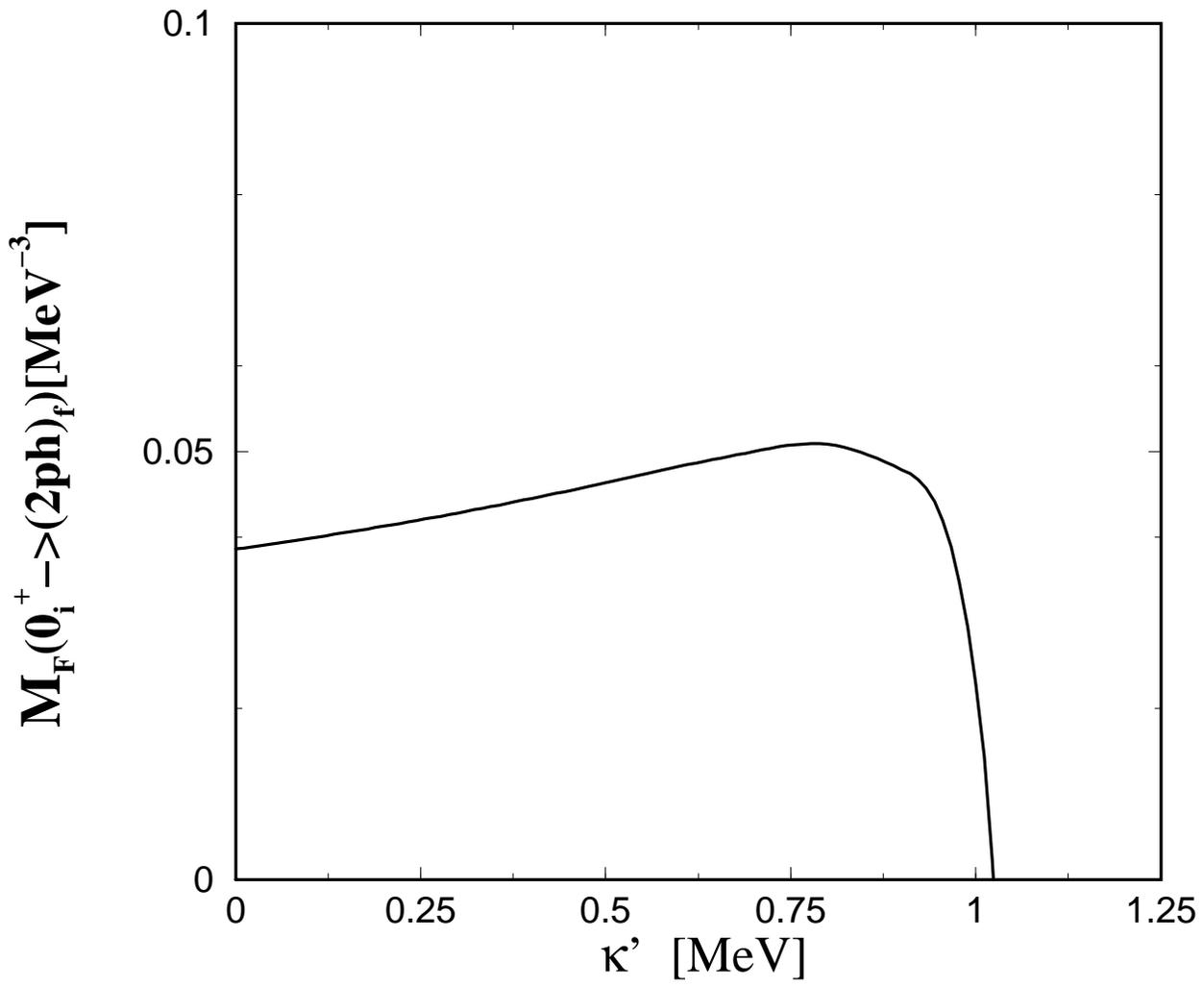}
\caption{The transition amplitude from the ground state of the mother
 nucleus to the two pairing vibrational states, characterizing the daughter 
nucleus, is plotted as a function of the proton-neutron pairing strength k'.}
\end{figure}
\clearpage

FIGURE 7

\begin{figure}[ht!]
\vskip 20.2cm 
\includegraphics{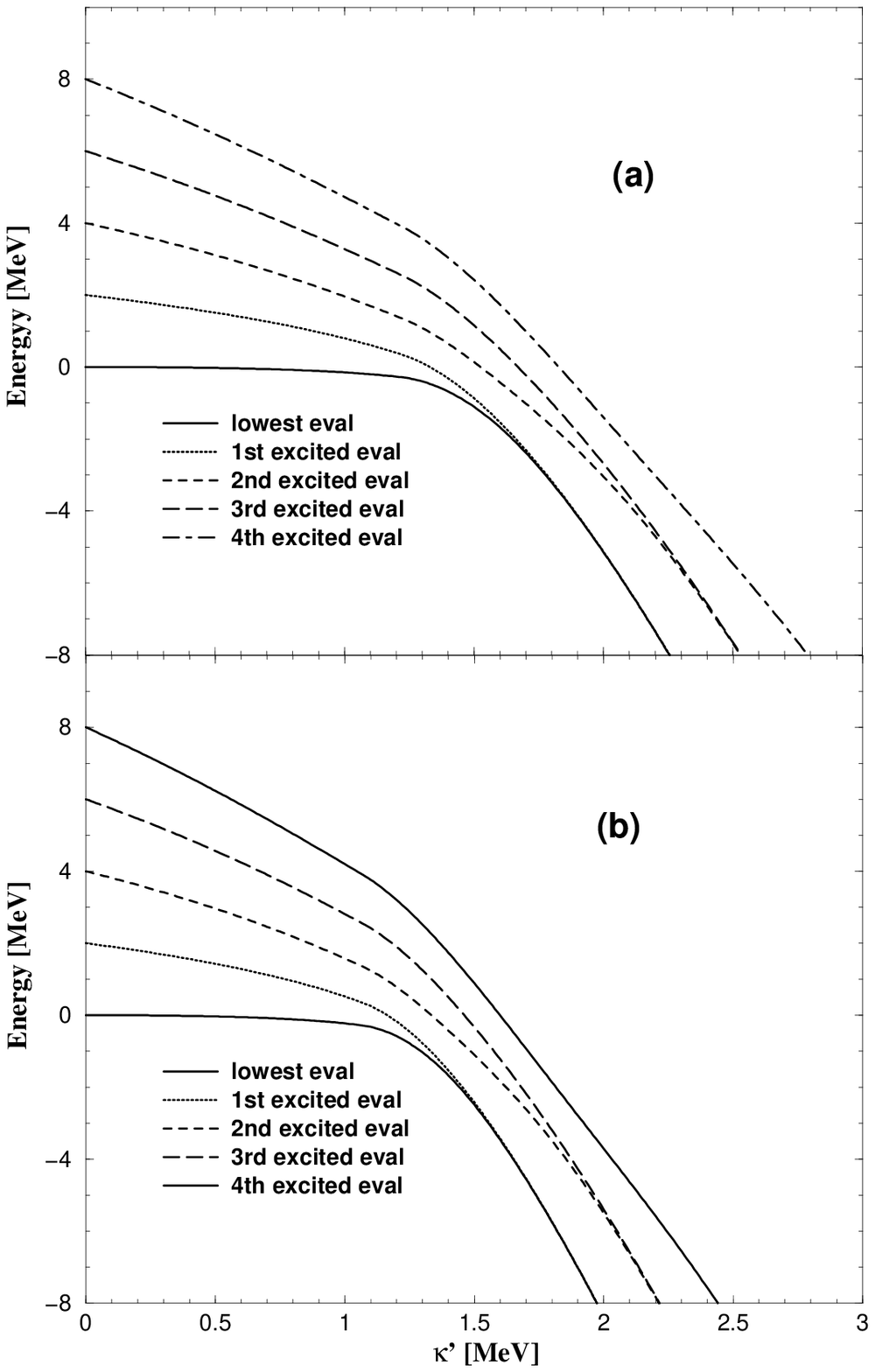}
\caption{The first five eigenvalues of the Holstein-Primakoff boson expanded
 Hamiltonian (4.4), for mother (a) and daughter (b) nuclei, are represented 
as functions of k'. }
\end{figure}
\clearpage

FIGURE 8
\begin{figure}[ht!]
\vskip 19cm 
\includegraphics{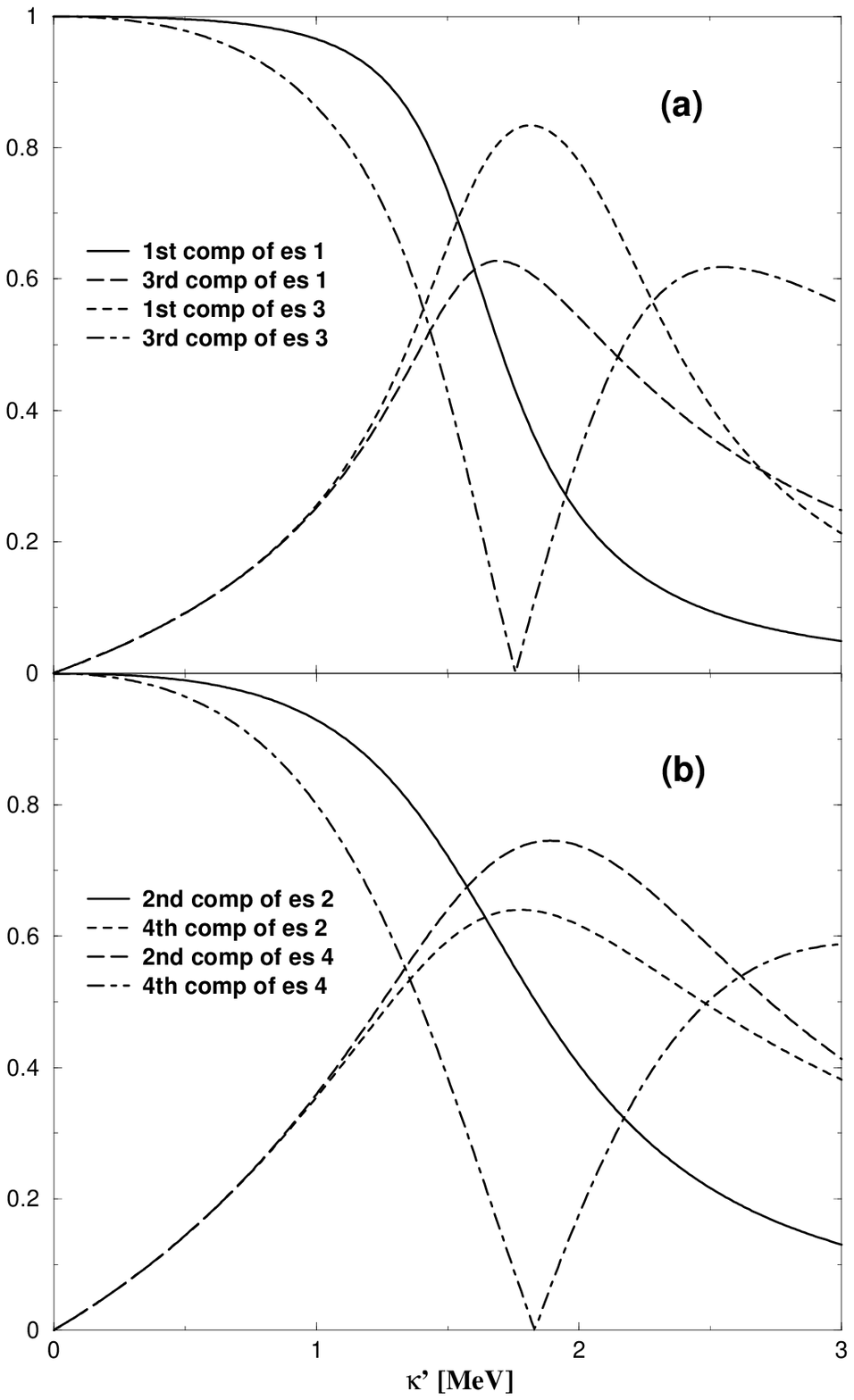}
\caption{The absolute values for the leading two amplitudes of the 
first four eigenstates of 
the Holstein-Primakoff boson expanded Hamiltonian given by Eq. (4.4) 
are represented as functions of k'.
In the upper panel the 1st and 3rd eigenstates are analyzed while in the 
lower panel the 2nd and the 4th are given.} 
\end{figure}
\clearpage

FIGURE 9
\begin{figure}[ht!]
\vskip 20.2cm 
\includegraphics{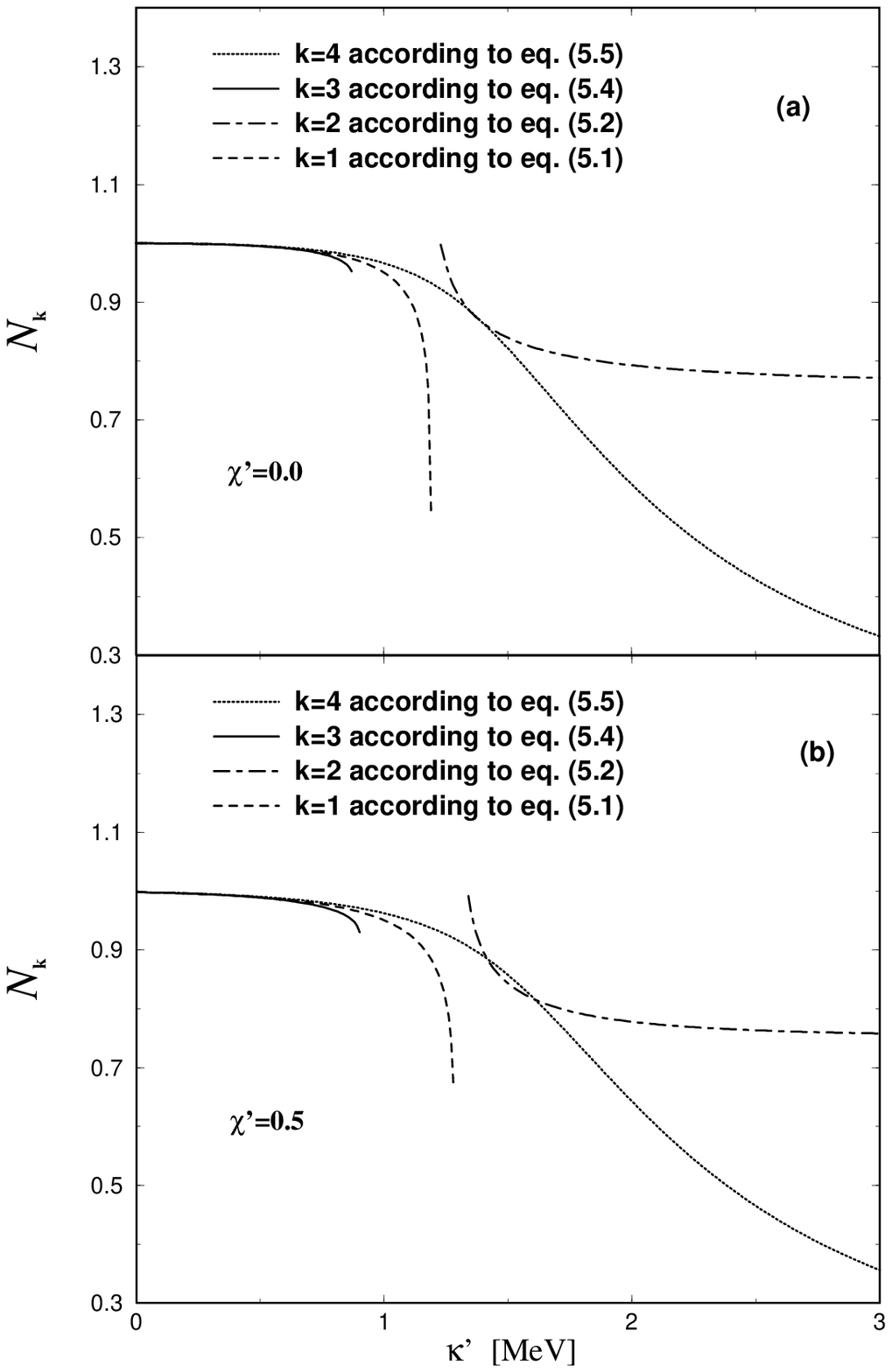}
\caption{ The norms of the states defined by eqs.. (5.1),(5.2),(5.4),(5.5)
are plotted as functions of the proton-neutron pairing strength k'.}
\end{figure}
\clearpage

FIGURE 10
\begin{figure}[ht!]
\vskip 20.2cm 
\includegraphics{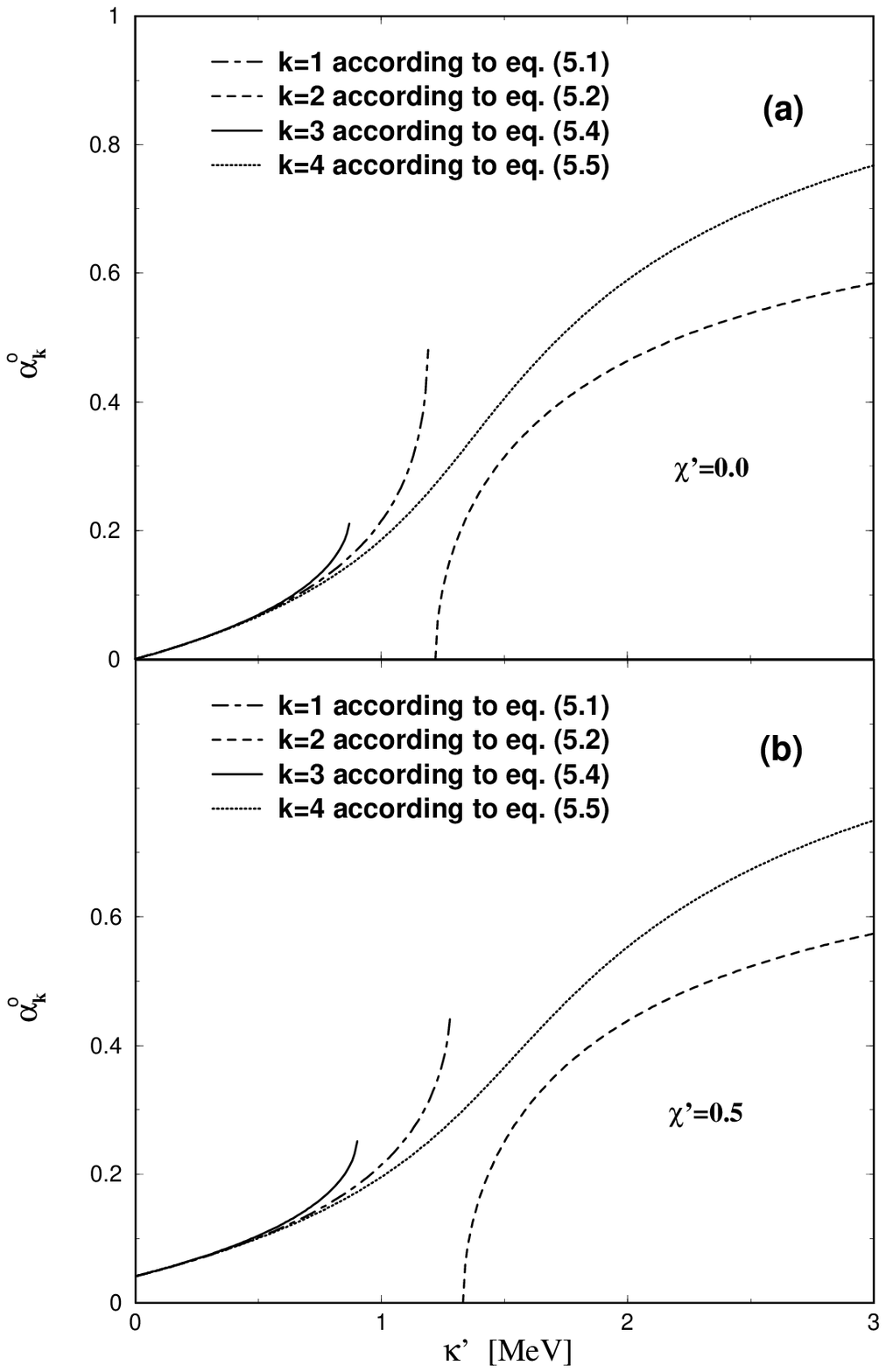}
\caption{The exponent $\stackrel{\circ}{\alpha}_k$
entering the 
definition of the  states $|\Phi_k\rangle$ given by Eq. (5.7).}
\end{figure}
\clearpage

FIGURE 11
\begin{figure}[ht!]
\vskip 20.2cm 
\includegraphics{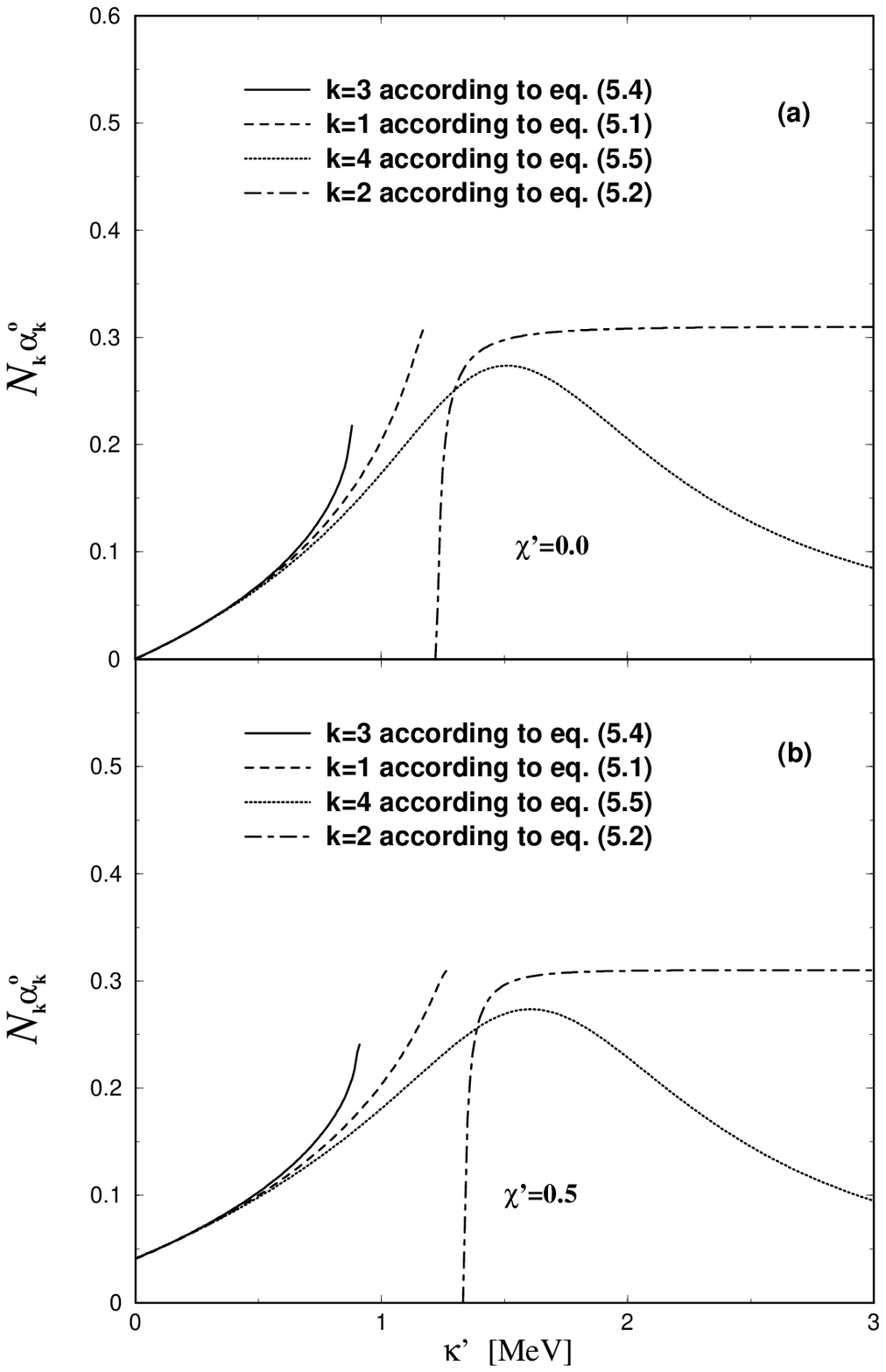}
\caption{The weights for a two phonon state in the structure of 
the states $|\Phi_k\rangle$ are plotted as function of the 
strength of the particle particle interaction.}
\end{figure}

\begin{references}
\bibitem{Hax}W. C. Haxton and G. J. Stephenson, Jr., Prog. Part. Nucl. Phys.
{\bf 12} (1984) 409.
\bibitem{Doi}M. Doi, T. Kotani and E. Takasugi, Prog. Theor. Phys., Suppl.
{\bf 83} (1985) 1.
\bibitem{Fas}A. Faessler, Prog. Part. Nucl. Phys. 21 (1988)
2139. 
\bibitem{Tom} T. Tomoda, Rep. Prog. Phys. {\bf 54}(1991) 53.
\bibitem{SuCi}J. Suhonen, O. Civitarese, Phys. Rep. 300 (1998) 2139.
\bibitem{Fas1}A. Faessler and F. \v {S}imkovic, J. Phys. G
{\bf24}(1998) 2139. 
\bibitem{vogel1}P. Vogel and P. Fischer, Phys. Rev. C {\bf32} (1985) 1362.
\bibitem{vogel2}P. Vogel and M. R. Zirnbauer, Phys. Rev. Lett. {\bf57}
(1986) 3148 .
\bibitem{civ1}O. Civitarese, A. Faessler and T. Tomoda, Phys. Lett. B 
{\bf194} (1986) 11.
\bibitem{grotz}K. Grotz, H. V. Klapdor and J. Metzinger, Phys. Lett. B
{\bf132} (1983) 22.
\bibitem{Klap}H. V. Klapdor and K. Grotz, Phys. Lett. B {\bf142} 
(1984) 323.
\bibitem{mut}K. Muto, E. Bender and H. V. Klapdor, Z. Phys. A {\bf334}, 
(1989) 177.
\bibitem{sutai}J. Suhonen, T. Taigel and A. Faessler, Nucl.
Phys.{\bf A486} (1988) 91.
\bibitem{rad1}A. A. Raduta, A. Faessler, S. Stoica and W. Kaminski, 
Phys. Lett. B {\bf 254} (1991) 7 .
\bibitem{rad2}A. A. Raduta, A. Faessler and S. Stoica, 
Nucl. Phys. A {\bf534} (1991) 149.
\bibitem{suh1}J. Suhonen, Nucl. Phys. A {\bf 563} (1993) 205 .
\bibitem{gri}A. Griffiths and P. Vogel, Phys. Rev. C {\bf46}
(1992) 181.
\bibitem{RadSuh1} A. A. Raduta and J. Suhonen, Phys. Rev. {\bf C53} (1996) 176.
\bibitem{RadSuh2} A. A. Raduta and J. Suhonen, J. Phys. {\bf G} N. Ph. {\bf 22}
(1996) 123.  
\bibitem{SimSmo}F. \v {S}imkovic, A. A. Raduta, M. Veselsky, A.
Faessler, Phys. Rev. {\bf C 61} (2000) 044319.
\bibitem{Toi95}J. Toivanen, J. Suhonen, Phys. Rev. Lett.{\bf 75}
(1995) 410. 
\bibitem{Schw96}J. Schwieger, F. \v {S}imkovic and A. Faessler, 
Nucl. Phys.{\bf A600} (1996) 179.
\bibitem{SimKam}J. Schwieger F. \v {S}imkovic, A. Faessler
and W. A. Kaminski, J. Phys. G {\bf 23} (1997) 1647; Phys. Rev.
{\bf C57} (1998) 1738.
\bibitem{RadRad}A.A. Raduta, M.C. Raduta, A. Faessler,W. Kaminski, 
Nucl. Phys.{\bf A 634}(1998) 497.
\bibitem{Rasifa}A. A. Raduta, F. \v {S}imkovic, A. Faessler, Jour.
Phys. {\bf G 26}(2000) 793.
\bibitem{RaHa}A. A. Raduta, O. Haug, F. \v {S}imkovic, A. Faessler,
Nucl. Phys. {\bf A 671} (2000) 255.
\bibitem{RaHa2} A. A. Raduta, O. Haug, F. \v {S}imkovic, A. Faessler,
Jour. Phys. {\bf G 26} (2000) 1327.
\bibitem{RaPa} A. A. Raduta, L. Pacearescu, V. Baran,
P.Sariguren and E. Moya de Guerra, Nucl. Phys. {\bf A 675}
(2000) 503.
\bibitem{Samb} M. Sambataro and J. Suhonen, Phys. Rev. {\bf C56} (1997) 782.
\bibitem{Civ1} J. G. Hirsch, P. O. Hess and O. Civitarese, Phys. Rev. {\bf C 56}
(1997) 199.
\bibitem{Civ2} O. Civitarese, P. O. Hess, J. G. Hirsch and M. Reboiro,
{\bf C 59} (1999) 194.
\bibitem{Rad11} A. A. Raduta, D. S. Delion and Amand Faessler, Phys. Rev.
{\bf C 51} (1995) 3008.
\bibitem{Rad12} A. A. Raduta, D. S. Delion and Amand Faessler, Nucl. Phys. 
{\bf A 617} (1997) 176.
\bibitem{Hols}T. Holstein and H. Primakoff, Phys. Rev {\bf 58}
(1940) 1098.
\bibitem{Schw}J. Schwinger "On angular momentum" in Quantum
Theory of Angular Momentum, 1965, edited by L. Biedenharn H. Van
Dam (Academic, New York), p. 229.
\end{references}
\end{document}